\documentclass[aps,prx,10pt,twocolumn,superscriptaddress,longbibliography]{revtex4-2}

\usepackage{graphicx}
\usepackage{svg}
\usepackage{dcolumn}
\usepackage{bm}
\usepackage{amsmath}
\usepackage{amsthm}
\usepackage{amssymb}
\usepackage{mathrsfs}
\usepackage[colorlinks,
            linkcolor=black,
            citecolor=black,
            urlcolor=blue
            ]{hyperref}
\usepackage{tabularx} 
\usepackage{natbib}
\usepackage[utf8]{inputenc}
\usepackage{booktabs} 
\usepackage{multirow} 
\usepackage{siunitx}  
\usepackage{xcolor}   
\newcommand{\makecell}[2][]{\shortstack{#2}}
\usepackage{cleveref}

\newcommand{\be}{\begin{equation}}
\newcommand{\ee}{\end{equation}}

\begin{document}
\title{Nonequilibrium Dynamics of Simple Exclusion Processes Across Dimensions}


\author{Zhimao Liu}
\affiliation{Institute of Fundamental and Frontier Sciences, University of Electronic Science and Technology of China, Chengdu 611731, China}


\author{Jing Liu}
\affiliation{School of Physical Science and Technology, Beijing University of Posts and Telecommunications, Beijing 100876, China}
\affiliation{Institute of Theoretical Physics, Chinese Academy of Sciences, Beijing 100190, China}

\author{Pan Zhang}
\email[Corresponding authors: ]{panzhang@itp.ac.cn}
\affiliation{Institute of Theoretical Physics, Chinese Academy of Sciences, Beijing 100190, China}
\affiliation{School of Fundamental Physics and Mathematical Sciences, Hangzhou Institute for Advanced Study, UCAS, Hangzhou 310024, China}

\author{Ying Tang}
\email[Corresponding authors: ]{jamestang23@gmail.com}
\affiliation{Institute of Fundamental and Frontier Sciences, University of Electronic Science and Technology of China, Chengdu 611731, China}
\affiliation{School of Physics, University of Electronic Science and Technology of China, Chengdu 611731, China}
\affiliation{Key Laboratory of Quantum Physics and Photonic Quantum Information, Ministry of Education, University of Electronic Science and Technology of China, Chengdu 611731, China}
\affiliation{Non-classical Information Science Basic Discipline Research Center of Sichuan Province, University of Electronic Science and Technology of China, Chengdu 611731, China}





\begin{abstract}
The simple exclusion process (SEP) is a paradigmatic model for nonequilibrium transport, yet its rich dynamics over an exponentially large configuration space remain notoriously intractable. Here, we leverage variational autoregressive networks to characterize the nonequilibrium dynamics of the symmetric (SSEP), asymmetric (ASEP), and totally asymmetric (TASEP) cases in one to three dimensions at any time.  We validate the approach against established finite-time 1D and long-time 2D results for the SSEP, and further provide new results on dynamical activity, density fields and nonequilibrium phase transitions for 2D and 3D cases. In 2D, we give a directional-density criterion
connecting the bulk-density organization to the 1D TASEP phase diagram, and reveal how boundary and bulk rates separately control the activity and susceptibility. In 3D, we provide the first finite-time characterization of the SSEP active-inactive phase transition, and uncover scaling relations over time. Across dimensions, the critical field follows a trend $s_c\sim L^{-2}$, suggesting tuning strategies by the characteristic diffusive length scale. Overall, this work establishes a unified neural-network framework for characterizing nonequilibrium dynamics of representative transport systems.
\end{abstract}

\maketitle


\section{\label{sec:Introduction}
Introduction
}
The simple exclusion process (SEP) is a paradigmatic model of nonequilibrium transport in which particles hop stochastically on a lattice subject to hard-core exclusion~\cite{28_spitzer1970interaction,27_schmittmann1995statistical}. Its symmetric (SSEP), asymmetric (ASEP), and totally asymmetric (TASEP) variants range between diffusion and ballistic transport. Despite their minimal microscopic rules, boundary reservoirs and biased hopping generate rich steady-state density profiles~\cite{31_derrida1993exact,1_derrida2007non,10_gorissen2012exact}, boundary-induced phase transitions~\cite{31_Derrida1992,31_derrida1993exact,schutz1993phase}, density shocks~\cite{79_derrida1997shock}, and current fluctuations~\cite{11_derrida1998exact,87_bertini2005current,86_appert2008universal}. These properties make SEP a common framework for queuing dynamics~\cite{78_ha2003queuing}, molecular-motor traffic~\cite{108_miedema2017correlation}, vehicular traffic flow~\cite{6_schadschneider2000statistical}, kinetic surface growth~\cite{4_krug1997origins}, and polymer transport~\cite{29_widom1991repton}. Its broader connections to biological transport and driven many-body physics are reviewed in~\cite{2_Chou_2011}. At the trajectory level~\cite{100_pagare2024stochastic,109_stutzer2026stochastic}, integrated current measures net directed particle transport~\cite{87_bertini2005current}, whereas dynamical activity counts all particle jumps, including injection and removal at the boundaries~\cite{85_lecomte2007thermodynamic}; their fluctuations probe complementary time-antisymmetric and time-symmetric sectors of nonequilibrium dynamics~\cite{80_maes2020frenesy,82_vanicat2021mapping,gorissen2009density}. A complete dynamical description therefore requires not only typical densities or currents but also the fluctuations of these time-integrated observables.

Analytical understanding of exclusion processes is more complete in one dimension and in the steady-state or long-time limits. The matrix product ansatz (MPA) provides exact stationary states and boundary-induced phase diagrams for open systems~\cite{31_Derrida1992,31_derrida1993exact,110_DERRIDA199865}. Bethe-ansatz and exact master-equation studies further characterize one-dimensional dynamics and spectral properties~\cite{14_golinelli2006asymmetric,schutz1997exact}, while exact and finite-size analyses establish stationary density and current large deviations~\cite{11_derrida1998exact,32_derrida2003exact,50_de2011large,51_gorissen2011finite,10_gorissen2012exact}. More generally, fluctuations of time-integrated observables are analyzed by large-deviation theory~\cite{43_donsker1975asymptotic,8_touchette2009large} and the thermodynamic formalism for Markov dynamics~\cite{85_lecomte2007thermodynamic,lecomte2007continuous}, in which a tilted generator reweights trajectories according to the chosen observable. In the long-time limit, the dominant eigenvalue of this generator gives the scaled cumulant generating function (SCGF)~\cite{9_garrahan2009first}, while macroscopic fluctuation theory provides a complementary coarse-grained description of rare trajectories~\cite{84_Bertini2015Macroscopic}. This long-time framework has revealed, for example, that suppressing the activity of the SSEP can induce an inhomogeneous clustered phase~\cite{81_lecomte2012inactive}. At finite time, however, the dynamical partition function $Z_t(s)$ depends on the initial state and receives contributions from the full spectrum of the tilted generator~\cite{12_causer2022finite}. Finite-time characterization therefore requires tracking the evolution of a distribution over an exponentially large configuration space while retaining multiple transient modes, a task that becomes increasingly difficult with system size and spatial dimension.

The finite-time and high-dimensional challenges have motivated diverse numerical strategies for sampling or representing biased trajectory ensembles. Importance sampling~\cite{13_ray2018importance,41_ray2018exact}, umbrella sampling~\cite{47_klymko2018rare,42_perez2019sampling}, population dynamics~\cite{45_nemoto2017finite,54_brewer2018efficient}, and variational rare-event methods~\cite{48_jacobson2019direct,73_rose2021reinforcement} sample rare trajectories, although their accuracy and efficiency can depend on auxiliary dynamics or population-control effects. Representation-based approaches instead compress the biased distribution: tensor networks enable leading-eigenstate and rare-trajectory calculations in one dimension~\cite{68_banuls2019using,69_causer2021optimal}, explicit finite-time evolution~\cite{12_causer2022finite}, and long-time studies of two-dimensional dynamical ensembles~\cite{18_helms2020dynamical,19_causer2023optimal}. Neural networks and autoregressive models provide another compact representation of many-body distributions, with applications to classical statistical mechanics~\cite{16_wu2019solving,20_Liu2025Efficient,zhong2026scalable}, stochastic reaction networks~\cite{tang2023neural,93_fang2023divide,92_Chuanbo_2024Distilling,weng2025tracking,99_cai2026revival}, many-particle dynamics~\cite{104_xiong2025capturing,105_suzuki2025machine,106_zhu2026two}, nonequilibrium dynamical phase transitions~\cite{65_garrahan2007dynamical,23_casert2021dynamical,17_tang2024learning,102_zhou2024k,103_muzzi2024principal,94_liu2025dynamical}, quantum many-body systems~\cite{15_carleo2017solving,RevModPhys.91.045002,63_hibat2020recurrent,95_Melko2024Language,101_bulgarelli2025flow}, and open-system dynamics~\cite{22_luo2022autoregressive}. Despite these advances, finite-time evolution remains more demanding than stationary or long-time optimization because the evolving tilted distribution must be represented at successive times while multiple transient modes remain relevant. Consequently, controlled finite-time calculations are most developed in one dimension~\cite{12_causer2022finite}, while higher-dimensional studies focused on selected models, system sizes, or long-time observables~\cite{18_helms2020dynamical,23_casert2021dynamical,19_causer2023optimal,17_tang2024learning}. A systematic treatment of the finite-time biased dynamics of SSEP, ASEP, and TASEP from one to three dimensions remains absent.
\begin{figure*}[htbp]
\includegraphics[width=1\textwidth]{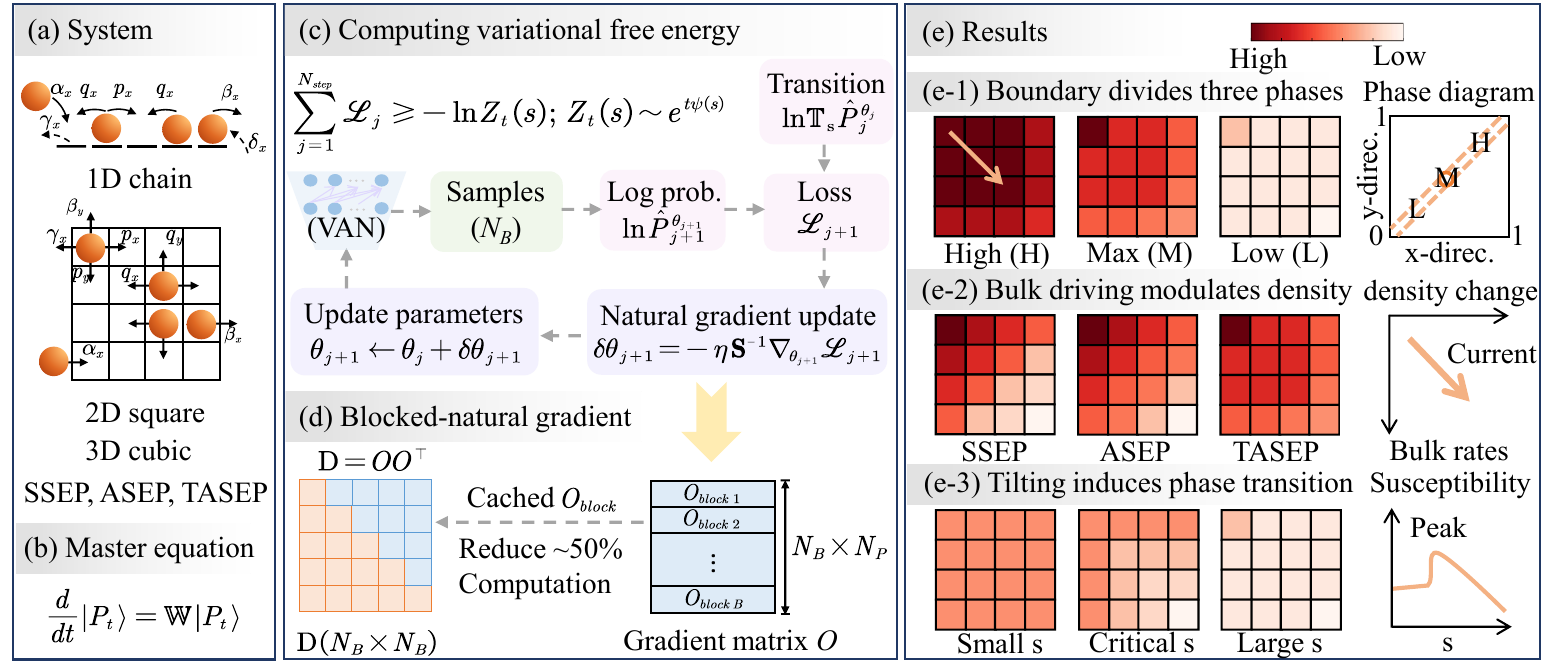}
\caption{\label{fig:main}\textbf{Unified framework for nonequilibrium dynamics of simple exclusion processes (SEPs) across dimensions.} (a) Illustration of SEP models on 1D chain, 2D square, and 3D cubic lattices, with the mathematical symbols denoting boundary and bulk rates. (b) Master equation (Eq.~\eqref{eq:1}) governing the time evolution of the probability vector $|P_t\rangle$ under the tilted generator $\mathbb{W}_s$ of symmetric (SSEP), asymmetric (ASEP), and totally asymmetric (TASEP) simple exclusion processes. (c) Pipeline for calculating the variational free energy $-\ln Z_t(s)$ (Eq.~\eqref{eq:2}) and the scaled cumulant generating function (SCGF) $\psi(s)$ (Eq.~\eqref{eq:4}) using a variational autoregressive network (VAN) with parameters $\theta_j$ at each time step, optimized via natural-gradient descent. (d) Blocked natural-gradient scheme partitioning the gradient matrix $O$ to reduce the computational cost of the $\mathrm{D}$ matrix. (e) Representative dynamical properties: (e-1) diagonal phase organization inferred from the directional-density criterion, where matched directional boundary densities identify low-density (L), maximal-current (M), and high-density (H) regimes; 
(e-2) bulk driving from SSEP to TASEP enhances the activity and susceptibility response via density modulation in the maximal-current sector; and (e-3) phase transitions induced by the tilting $s$, characterized by a peak in the susceptibility. 
}
\end{figure*}
\begin{figure*}[htbp]
\includegraphics[width=1\textwidth]{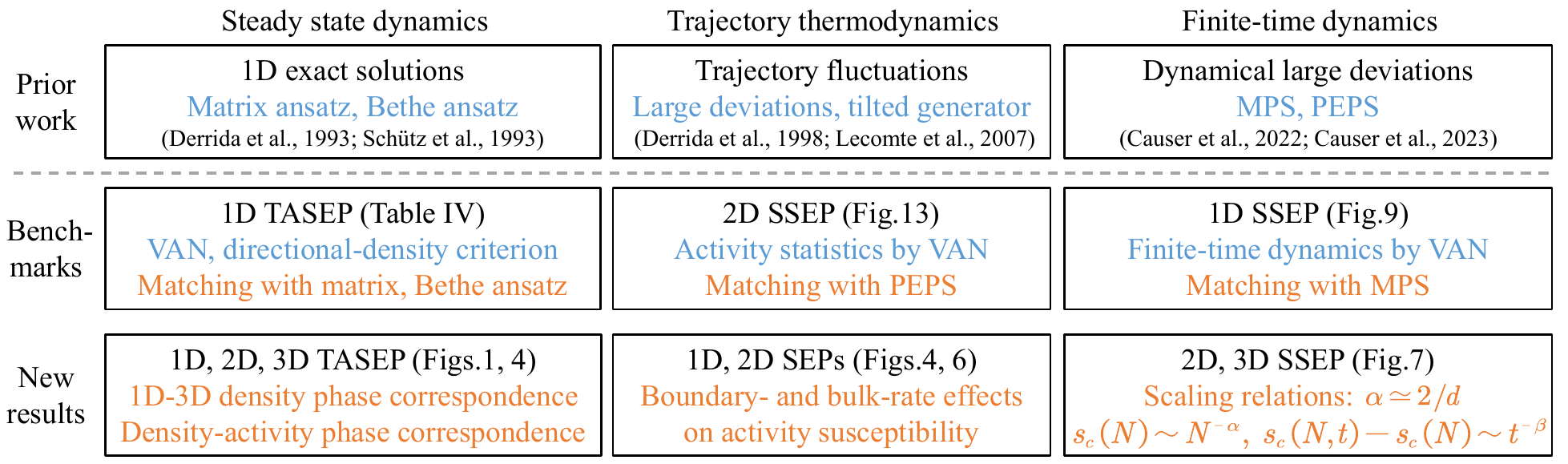}
\caption{\label{fig:pipeline}\textbf{Historical development and new results for exclusion-process dynamics.}
The upper row traces the progression from exact one-dimensional solutions based on matrix-product and Bethe-ansatz methods~\cite{31_derrida1993exact,schutz1993phase}, trajectory fluctuations and large-deviation theory~\cite{11_derrida1998exact,lecomte2007continuous}, and tensor-network approaches~\cite{12_causer2022finite,19_causer2023optimal}. The middle row shows results reproducing previous findings, including the 1D TASEP steady state, the dynamical activity of the 2D SSEP, and finite-time 1D SSEP dynamics. The lower row summarizes new results, including a density-activity phase correspondence and, through the directional-density criterion, a three-phase correspondence between the steady-state densities of 1D to 3D TASEP. It also shows the effects of boundary and bulk rates on the activity susceptibility and new finite-time scaling relations in 2D and 3D.
Black text denotes the physical systems or problems, blue text denotes methods and theoretical frameworks, and orange text are benchmarks and new results of this work. MPS and PEPS denote matrix product states and projected entangled-pair states, respectively.
}
\end{figure*}

To address this challenge, we build on the finite-time variational autoregressive network (VAN) framework~\cite{17_tang2024learning}, incorporate blocked natural-gradient optimization~\cite{20_Liu2025Efficient}, and systematically extend its application to SEP dynamics from one to three dimensions (Fig.~\ref{fig:main}). The VAN~\cite{16_wu2019solving,17_tang2024learning} represents normalized configuration probabilities and approximates the evolution of the tilted distribution variationally, allowing the dynamical partition function and its derived observables to be evaluated without full configuration-space enumeration. We apply this extended framework to SSEP, ASEP, and TASEP under different boundary conditions and bulk asymmetries. We first compare unbiased, long-time density and current estimates with exact unbiased 1D TASEP results (Appendix~\ref{app:1d-tasep-steady-state}), and then validate the finite-time dynamics against established 1D SSEP results~\cite{12_causer2022finite} (Appendix~\ref{app:1D SSEP benchmark}),
long-time tensor-network benchmarks for the 2D SSEP~\cite{19_causer2023optimal}, and exact numerical results for small systems. These benchmarks establish the accuracy required to examine transient regimes and higher-dimensional systems that are inaccessible to direct enumeration. Beyond accuracy, the computational scaling benchmarks in Fig.~\ref{fig:scaling} demonstrate access to 2D lattices with $L=10$ ($N=100$) on a single 80GB GPU. Extrapolating the measured Transformer memory scaling projects access to $N\approx225$-$361$ on four to eight such GPUs (Sec.~\ref{subsection:Blocked natural-gradient descent}), providing a concrete route to larger-system studies.

Our results are organized around dynamical activity and its response. Boundary rates and bulk driving shape the density profiles and thereby the activity, whereas the dynamical susceptibility distinguishes weakly and strongly responsive regimes and signals potential phase transitions. The 1D calculations provide a reference by connecting finite-time activity maps to the classical three-phase TASEP organization~\cite{31_Derrida1992,31_derrida1993exact}. The main new results concern higher dimensions. In 2D, we establish a directional-density criterion: when the boundary parameters along both directions select the same density according to the 1D TASEP phase diagram, the bulk density converges to this common value in the thermodynamic limit. We further show how boundary and bulk driving separately govern the 2D activity and susceptibility. In 3D, we obtain new finite-time and finite-size scaling relations for the SSEP. Comparing dimensions, the critical counting field exhibits a trend broadly consistent with $s_c\sim N^{-\alpha}$ and $\alpha\approx 2/d$, equivalent to $s_c\sim L^{-2}$ for $N=L^d$. This trend suggests that the active-inactive transition is asymptotically governed by the system's characteristic length scale. Together, these results provide a unified, activity-centered characterization of finite-time and long-time SEP dynamics across dimensions; a schematic summary is given in Fig.~\ref{fig:pipeline}.

The paper is organized as follows. In Sec.~\ref{sec:Nonequilibrium statistical mechanics}, we define the multidimensional exclusion processes and formulate the tilted-generator description of time-integrated dynamical observables. In Sec.~\ref{sec:Neural-network Framework}, we introduce the VAN ansatz, time-discretized variational evolution, and blocked natural-gradient optimization~\cite{24_amari1998natural,20_Liu2025Efficient}. Section~\ref{sec:Applications} applies this framework to boundary- and bulk-driven dynamics in 1D, directional density and dynamical phase behavior in 2D, and new scaling relations of the active-inactive transition in 3D. Finally, Sec.~\ref{sec:Discussion} summarizes the physical conclusions and numerical validation, discusses computational details, and outlines extensions to larger and more nonequilibrium systems. The notation and conventions used throughout the paper are summarized in Appendix~\ref{app:notation}.

\section{\label{sec:Nonequilibrium statistical mechanics}
Nonequilibrium statistical mechanics
}

\subsection{\label{sec:Master equations}Master equations}

We consider a continuous-time Markov process on a $d$-dimensional lattice with $N$ sites. Each site $i$ is characterized by a binary variable $x_i \in \{0, 1\}$, where $x_i=1$ denotes an occupied site and $x_i=0$ an empty one. A configuration of the system is represented by the vector $|\mathbf{x}\rangle = |x_1, x_2, \dots, x_N\rangle$. The probability distribution $P(\mathbf{x}, t)$ of the system in configuration $\mathbf{x}$ at time $t$ is encoded in the vector $|P_t\rangle = \sum_{\mathbf{x}} P(\mathbf{x}, t) |\mathbf{x}\rangle$. The time evolution of $|P_t\rangle$ is governed by a master equation:
\begin{equation}\label{eq:1}
    \frac{\partial}{\partial t} |P_t\rangle = \mathbb{W} |P_t\rangle, 
\end{equation}
where $\mathbb{W}$ is the Markov generator. The generator can be decomposed as $\mathbb{W} = \mathbb{K} - \mathbb{R}$, where $\mathbb{K}$ is an off-diagonal matrix containing 
transition rates $w_{\mathbf{x}\to\mathbf{y}}$ from configuration $\mathbf{x}$ to $\mathbf{y}$, with elements $\mathbb{K}_{\mathbf{y}\mathbf{x}}=w_{\mathbf{x}\to\mathbf{y}}$, and $\mathbb{R}$ is a diagonal matrix of escape rates $R_{\mathbf{x}} = \sum_{\mathbf{y} \neq \mathbf{x}} w_{\mathbf{x} \to \mathbf{y}}$, ensuring probability conservation $\langle - | \mathbb{W} = 0$ with $\langle - | = \sum_{\mathbf{x}} \langle \mathbf{x} |$.

\subsection{\label{sec:Dynamical partition function}Dynamical partition function}

To characterize the dynamical behavior of the system, we examine the ensemble of all possible trajectories $\omega_t = \{\mathbf{x}_{t_0} \to \mathbf{x}_{t_1} \to \dots \to \mathbf{x}_t\}$ in a time-discretized representation with step length $\delta t$, total time $t=N_{\mathrm{step}}\delta t$, and $N_{\mathrm{step}}$ time steps. Let $\hat{K}(\omega_t)$ be a time-extensive dynamical observable, such as the total number of configuration changes, i.e., the dynamical activity. This jump count is invariant under reversal of the trajectory and therefore probes the time-symmetric frenetic sector of the path ensemble~\cite{80_maes2020frenesy}. The probability of observing a specific value $K$ is given by $P_t(K) = \sum_{\omega_t} p(\omega_t) \delta(\hat{K}(\omega_t) - K)$, where $p(\omega_t)$ is the weight of the trajectory. The statistics of $K$ are captured by the moment-generating function, or dynamical partition function:
\begin{equation}\label{eq:2}
    Z_t(s) = \sum_K P_t(K) e^{-sK} = \langle - | e^{t\mathbb{W}_s} | P_0 \rangle,
\end{equation}
where $s$ is the counting field conjugate to $K$, and $\mathbb{W}_s$ is the ``tilted'' generator. For the dynamical activity, where each configuration change contributes one unit to $K$, the tilted generator takes the form $\mathbb{W}_s = e^{-s}\mathbb{K} - \mathbb{R}$. Unlike the original Markov generator, $\mathbb{W}_s$ does not preserve probability, and its largest eigenvalue determines the long-time behavior. At arbitrary finite times $t$ within the simulated time window, however, the partition function $Z_t(s)$ depends on the full spectrum of $\mathbb{W}_s$. This spectral dependence distinguishes the finite-time regime from the steady state because sub-leading decaying modes contribute to transient dynamical correlations~\cite{12_causer2022finite,17_tang2024learning,lin2026dynamical}.

At finite trajectory time $t$, we define the finite-time scaled cumulant generating function (SCGF), dynamical activity, and dynamical susceptibility as
\begin{equation}\label{eq:3}
 \begin{aligned}
    \psi_t(s) = \frac{1}{t}\ln Z_t(s),
    k_t(s) = -\frac{1}{N}\frac{\partial \psi_t(s)}{\partial s},
    \chi_t(s) = \frac{\partial^2\psi_t(s)}{\partial s^2}.
 \end{aligned}
\end{equation}
Here, $k_t(s)$ is the biased-ensemble dynamical activity per unit time and per site, whereas $\chi_t(s)$ is the extensive dynamical susceptibility. Numerically, $\chi_t(s)$ is evaluated using a centered three-point finite-difference scheme adapted to the nonuniform counting-field grid. The finite-size transition point $s_c(N,t)$ is then identified using the susceptibility-peak criterion.

In the long-time limit ($t \to \infty$), the dynamical partition function follows the large-deviation form $Z_t(s)\sim e^{t\psi(s)}$. The long-time SCGF is therefore
\begin{equation}\label{eq:4}
    \psi(s)=\lim_{t\to\infty}\psi_t(s)
    =\lim_{t\to\infty}\frac{1}{t}\ln Z_t(s).
\end{equation}
According to the Perron-Frobenius theorem, $\psi(s)$ is the dominant eigenvalue, i.e., the eigenvalue with the largest real part, of the tilted generator $\mathbb{W}_s$. The corresponding long-time activity and susceptibility are
\begin{equation}\label{eq:5}
    \begin{aligned}
        k(s)
        =-\frac{1}{N}\psi'(s),
        \chi(s)=\psi''(s).
    \end{aligned}
\end{equation}
At $s=0$, $k(0)$ is the typical steady-state activity per site and $\chi(0)$ is the steady-state growth rate of the activity variance. More generally, derivatives of $\psi(s)$ at $s=0$, with the alternating signs implied by the factor $e^{-sK}$ in Eq.~\eqref{eq:2}, yield the steady-state cumulants of $K$. A non-analyticity in $\psi(s)$ signifies a dynamical phase transition, marking a singular change in the trajectory ensemble's properties, such as a transition from an active phase to an inactive phase~\cite{8_touchette2009large,9_garrahan2009first,81_lecomte2012inactive}.

\subsection{\label{subsec:model}System}

The asymmetric simple exclusion process (ASEP) is a paradigmatic model of nonequilibrium transport, defined as a continuous-time Markov process on a lattice where ``hard-core'' particles hop stochastically between sites. The fundamental constraint is the exclusion principle: each site $i$ can be occupied by at most one particle, as encoded by the configuration variable $x_i \in \{0, 1\}$. The dynamics are governed by a master equation where the configuration $|\mathbf{x}\rangle$ evolves via local hopping events and boundary reservoirs. We consider the ASEP in one, two, and three dimensions, each defined by specific lattice geometries and anisotropic rates. As shown schematically in Fig.~\ref{fig:main}(a), the geometries are initialized from a spatially uniform probability distribution, denoted by $|P_0\rangle$, which serves as the unbiased initial ensemble before the tilted evolution is applied and is represented by the same orange color in Fig.~\ref{fig:main}(a).

The 1D ASEP takes place on a linear chain with $N=L$ sites. Each site may be occupied by a particle or empty. Particles stochastically hop into vacant nearest-neighbor sites in the right and left directions at rates $p_x$ and $q_x$, respectively. At the $\{$left; right$\}$ boundaries, particles are inserted at rates $\{\alpha_x; \delta_x\}$ and removed at rates $\{\gamma_x; \beta_x\}$.

The 2D ASEP takes place on a square lattice with $N=L^2$ sites. Each site may be occupied by a particle or empty. Particles stochastically hop into vacant nearest-neighbor lattice sites in the right (down) and left (up) directions at rates $p_x$ ($p_y$) and $q_x$ ($q_y$), respectively. At the $\{$left; top; right; bottom$\}$ boundaries, particles are inserted at rates $\{\alpha_x; \alpha_y; \delta_x; \delta_y\}$ and removed at rates $\{\gamma_x; \gamma_y; \beta_x; \beta_y\}$.

The 3D ASEP takes place on a cubic lattice with $N=L^3$ sites. Each site may be occupied by a particle or empty. Particles stochastically hop into vacant nearest-neighbor lattice sites in the right, down, and back directions at rates $p_x$, $p_y$, and $p_z$, and in the left, up, and front directions at rates $q_x$, $q_y$, and $q_z$, respectively. At the $\{$left; top; back; right; bottom; front$\}$ boundaries, particles are inserted at rates $\{\alpha_x; \alpha_y; \alpha_z; \delta_x; \delta_y; \delta_z\}$ and removed at rates $\{\gamma_x; \gamma_y; \gamma_z; \beta_x; \beta_y; \beta_z\}$.

We study the ASEP across various dimensions $d \in \{1, 2, 3\}$. The tilted generator is built from hopping, insertion and removal operators. For an oriented nearest-neighbor hop $i\to j$ along direction $\mu\in\{x,y,z\}$, the local tilted operator is
\begin{equation}\label{eq:6}
    o_{i\to j}^{\mathrm{hop}}
    =r_{i\to j}\left(e^{-s}a_i a_j^{\dagger}-n_i v_j\right),
\end{equation}
where $r_{i\to j}=p_\mu$ for a hop in the positive $\mu$ direction and $r_{i\to j}=q_\mu$ for the reverse direction. In the standard operator representation of stochastic dynamics~\cite{111_delRazo2026field,18_helms2020dynamical}, $a_i$ and $a_i^{\dagger}$ are the annihilation and creation operators, while $n_i=a_i^{\dagger}a_i$ and $v_i=\mathbb{I}-n_i$ are the particle-number and vacancy operators, respectively; in particular, $n_i|\mathbf{x}\rangle=x_i|\mathbf{x}\rangle$. For a boundary site $i$ coupled to a reservoir, the tilted insertion and removal operators are
\begin{equation}\label{eq:boundary-operators}
o_i^{\mathrm{in}}=r_i^{\mathrm{in}}\left(e^{-s}a_i^\dagger-v_i\right),
o_i^{\mathrm{out}}=r_i^{\mathrm{out}}\left(e^{-s}a_i-n_i\right),
\end{equation}
where $r_i^{\mathrm{in}}$ and $r_i^{\mathrm{out}}$ take the appropriate boundary rates $\alpha_\mu$, $\delta_\mu$, $\gamma_\mu$, or $\beta_\mu$ defined above.

\section{\label{sec:Neural-network Framework}Neural-network framework}
\subsection{Variational ansatz}
\begin{figure*}
\centering
\includegraphics[width=1\textwidth]{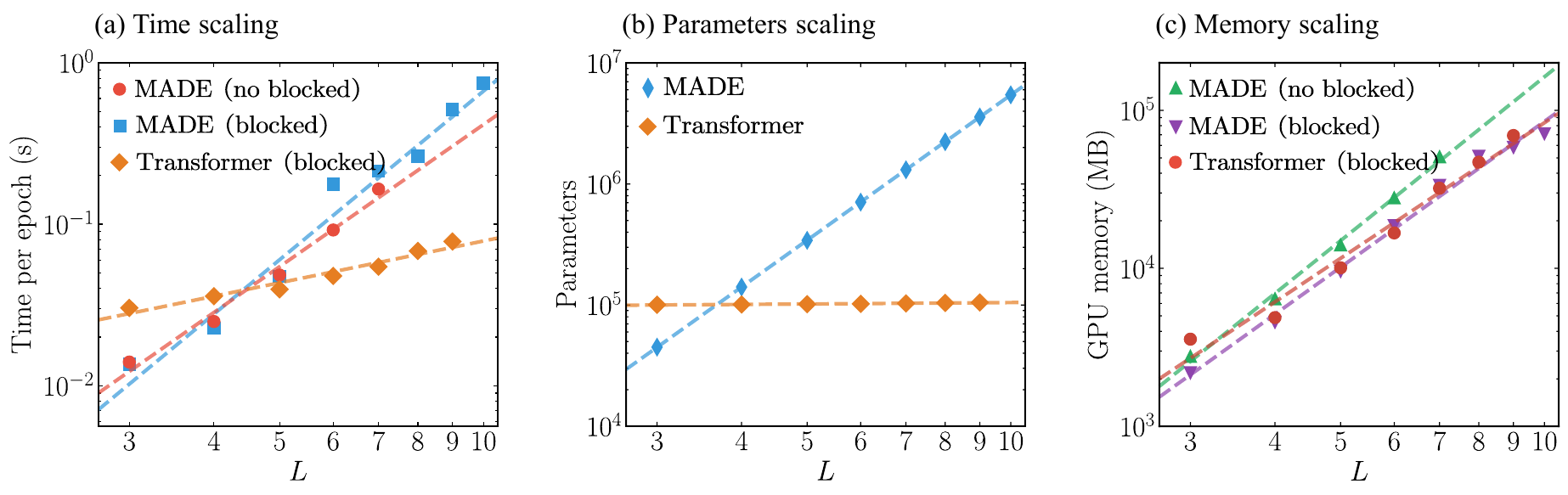}
\caption{\label{fig:scaling}\textbf{Computational time, parameter count, and GPU memory scaling with lattice size $L$ for the 2D systems.} 
(a) Time per epoch (in seconds) as a function of \(L\). For MADE, the computational time scales as \(L^{3.47}\) with the blocked natural-gradient scheme (blue squares) and as \(L^{2.92}\) without blocking (red circles), while the Transformer scales as \(L^{0.86}\) with blocked natural-gradient optimization (orange diamonds). (b) Scaling of the total number of network parameters with \(L\). The parameter count of MADE scales as \(L^{3.99}\) (blue diamonds), whereas that of the Transformer scales as \(L^{0.04}\) (orange diamonds). (c) GPU memory consumption as a function of \(L\). For MADE, GPU memory scales as \(L^{3.44}\) without blocking (green triangles) and as \(L^{3.07}\) with blocking (purple inverted triangles), while the Transformer scales as \(L^{2.85}\) (red circles).
}
\end{figure*}
To track the evolution equation, Eq.~\eqref{eq:1} or in Fig.~\ref{fig:main}(b), we use a neural-network
model, the VAN~\cite{16_wu2019solving,17_tang2024learning}, as a variational ansatz for the exact probability vector $|P_t\rangle$. The exact probability distribution is denoted by $P_t(\mathbf{x})$, while its neural-network approximation is denoted by $\hat{P}_{t}^{\theta}(\mathbf{x})$. The dependence of these distributions on the fixed counting field $s$ is left implicit unless displayed explicitly. The VAN represents this normalized distribution through an autoregressive factorization,
\begin{equation}\label{eq:7}
  \hat{P}_{t}^{\theta}(\mathbf{x}) = \prod_{i=1}^{N} \hat{P}_{t}^{\theta}(x_i | x_1, \dots, x_{i-1}),
\end{equation}
where the hat symbol denotes the parameterization by the neural
network with learnable parameters $\theta$. Here, we discretize time $t$ into intervals of size $\delta t$ and set $t_j=j\delta t$ for $j=0,\ldots,N_{\mathrm{step}}-1$. At each time step, the distribution $|\hat{P}_j^{\theta_{j}}\rangle$ is represented by an autoregressive neural architecture. We use a masked autoencoder for distribution estimation (MADE)~\cite{21_Germain2015MADE} for the smaller systems and a Transformer-based VAN~\cite{zhong2026scalable} for the 3D $L=5$ system; its architecture is specified in Appendix~\ref{app:Computational cost}.

For MADE, the autoregressive property, closely related to autoregressive neural-network wave-function constructions~\cite{16_wu2019solving,63_hibat2020recurrent}, is enforced by binary masking matrices, which ensure that $x_i$ depends only on predecessors $x_1, \dots, x_{i-1}$ and not on successors $x_{i+1}, \dots, x_N$~\cite{21_Germain2015MADE}. The Transformer enforces the same ordering through causal attention. Let $\mathbf{W}^l \in \mathscr{R}^{d^l \times d^{l-1}}$ be the weight matrix connecting layer $l-1$ to layer $l$, and let $\mathbf{M}^l \in \{0, 1\}^{d^l \times d^{l-1}}$ be the corresponding binary mask. This masking converts ordinary feed-forward propagation into an ordered conditional representation. The MADE forward propagation is defined as
\begin{equation}\label{eq:8}
    \mathbf{h}^l = \phi((\mathbf{W}^l \odot \mathbf{M}^l)\mathbf{h}^{l-1} + \mathbf{b}^l),
\end{equation}
where $\odot$ denotes the element-wise Hadamard product, $\phi$ is a non-linear activation function such as ReLU, $\mathbf{h}^0 = \mathbf{x}$ is the input layer, and $\mathbf{b}^l$ is the bias. The masks $\mathbf{M}^l$ are constructed such that a hidden unit $k$ in layer $l$ is connected to an input unit $x_i$ only if $m_{k,i}^{(l)} = 1$. The autoregressive condition is then satisfied by assigning each hidden unit $k$ a maximum order number $m(k)$ and enforcing $m_{k,i}^{(l)} = 0$ whenever $m(k) \ge i$. Thus, the ansatz preserves the causal ordering required for normalized sampling while remaining expressive enough for high-dimensional lattice distributions.

As summarized in Fig.~\ref{fig:main}(c), this variational evolution forms the computational pipeline used throughout the paper: the VAN represents the normalized probability distribution at each time step, the tilted operator advances the distribution in trajectory space, and the variational free energy provides the objective for updating the network parameters. Iterating this procedure yields the finite-time dynamical partition function \(Z_t(s)\), from which the SCGF \(\psi(s)\) and related dynamical observables are extracted.

Based on the VAN representation, we evaluate Eq.~\eqref{eq:1} with a counting field by applying
the operator $e^{\delta t \mathbb{W}_s}$ sequentially at each of the total $N_{\mathrm{step}}$ time steps: $Z_t(s)\approx \langle-|(\mathbb{T}_s)^{N_{\mathrm{step}}}|P_0\rangle$~\cite{17_tang2024learning}, where the transition operator $\mathbb{T}_{s} = \mathbb{I} + \delta t \mathbb{W}_s \approx e^{\delta t\mathbb{W}_s}$, and $\mathbb{I}$ denotes the identity operator. Without loss of generality, we consider the one-step evolution
from a normalized probability vector $|\hat{P}_{j}^{\theta_{j}}\rangle$ at time step $j$ for $j=0,\ldots,N_{\mathrm{step}}-1$. Evolution under $\mathbb{T}_{s}$ produces an unnormalized probability vector at each time step. Since the VAN represents
normalized probability distributions, the next variational state $|\hat{P}_{j+1}^{\theta_{j+1}}\rangle$ is optimized to approximate the normalized vector $\mathbb{T}_{s} |\hat{P}_{j}^{\theta_j}\rangle/Z_{j+1}(s)$. Here, $Z_{j+1}(s)$ is the one-step normalization factor for the update from $j$ to $j+1$, and the full dynamical partition function $Z_t(s)$ is obtained by accumulating these factors over all time steps. 

We perform this optimization by minimizing the Kullback-Leibler (KL) divergence~\cite{88_kullback1951information} between the VAN prediction at step $j+1$ and the normalized evolved distribution from step $j$. Equivalently, because the normalization factor $Z_{j+1}(s)$ is independent of $\theta_{j+1}$, the update minimizes the variational free energy $\mathscr{L}_{j+1}$~\cite{16_wu2019solving,17_tang2024learning}:
\begin{equation}\label{eq:9}
   \mathscr{L}_{j+1} = \mathbb{E}_{\mathbf{x} \sim \hat{P}_{j+1}^{\theta_{j+1}}} [\ln \hat{P}_{j+1}^{\theta_{j+1}}(\mathbf{x},s) - \ln (\mathbb{T}_{s} \hat{P}_{j}^{\theta_j}(\mathbf{x},s))].
\end{equation}
For the long-time estimates, we perform an independent log-log power-law extrapolation at each counting field by least-squares fitting the final five values of the negative log dynamical partition function. The extrapolated value is then used to evaluate the SCGF at the target long time. 

The gradient estimator follows directly from this variational free-energy objective and is related to broader variational optimization strategies for neural many-body states~\cite{25_chen2024empowering,77_Rende2024large-scale}. We use the score-function estimator REINFORCE~\cite{89_williams1992reinforce}, following its VAN implementation in~\cite{20_Liu2025Efficient}, and treat the term inside the expectation as the reward signal $\mathcal{R}(\mathbf{x})$:
\begin{equation}\label{eq:10}
    \mathcal{R}(\mathbf{x}) = \ln \hat{P}_{j+1}^{\theta_{j+1}}(\mathbf{x},s) - \ln (\mathbb{T}_{s} \hat{P}_{j}^{\theta_j}(\mathbf{x},s)).
\end{equation}
The gradient is then expressed as
\begin{equation}\label{eq:11}
   \nabla_{\theta_{j+1}} \mathscr{L}_{j+1} = \mathbb{E}_{\mathbf{x} \sim \hat{P}_{j+1}^{\theta_{j+1}}} [\mathcal{R}(\mathbf{x}) \nabla_{\theta_{j+1}} \ln \hat{P}_{j+1}^{\theta_{j+1}}(\mathbf{x},s)].
\end{equation}
For implementation, the expectation is approximated by drawing $N_B$ samples $\{\mathbf{x}^{(i)}\}_{i=1}^{N_B}$ from the current VAN~\cite{20_Liu2025Efficient}:
\begin{equation}\label{eq:12}
   \nabla_{\theta_{j+1}} \mathscr{L}_{j+1} \approx \frac{1}{N_B} \sum_{i=1}^{N_B} \mathcal{R}(\mathbf{x}^{(i)}) \nabla_{\theta_{j+1}} \ln \hat{P}_{j+1}^{\theta_{j+1}}(\mathbf{x}^{(i)}).
\end{equation}
Let $N_P$ be the number of trainable network parameters. By defining $O_{ik}=N_B^{-1/2}\partial\ln\hat{P}_{j+1}^{\theta_{j+1}}(\mathbf{x}^{(i)})/\partial \theta_k$ and $\mathcal{R}_i=N_B^{-1/2}\mathcal{R}(\mathbf{x}^{(i)})$, $O$ becomes an $N_B \times N_P$ scaled gradient matrix, and $\boldsymbol{\mathcal{R}}$ is an $N_B$-dimensional vector, Eq.~\eqref{eq:12} can be rewritten in matrix
notation as
\begin{equation}\label{eq:13}
   \nabla_{\theta_{j+1}} \mathscr{L}_{j+1} \approx O^{\top}\boldsymbol{\mathcal{R}}.
\end{equation}
This sampling form makes the update scalable, but the resulting high-dimensional parameter space can be ill-conditioned, so first-order optimizers such as Adam~\cite{90_kingma2015adam} may converge slowly~\cite{20_Liu2025Efficient}. A direct comparison in Appendix~\ref{app:optimizer-comparison} and Fig.~\ref{fig:Loss} confirms that, for the finite-time evolution considered here, natural-gradient descent requires substantially fewer optimization epochs and produces a smaller accumulated distribution error than Adam.
\subsection{Blocked natural-gradient descent}\label{subsection:Blocked natural-gradient descent}

Natural-gradient descent (NGD)~\cite{24_amari1998natural,20_Liu2025Efficient} addresses the ill-conditioning introduced by the high-dimensional autoregressive ansatz. Instead of measuring the update distance in Euclidean space, NGD measures the induced change in probability-distribution space. This geometry-aware update therefore minimizes the variational free energy under a fixed local KL-distance constraint, making the optimizer better matched to the probabilistic structure of the VAN than ordinary first-order methods.

The natural gradient update is defined as
\begin{equation}\label{eq:14}
    \delta \theta_{j+1} = -\eta \mathbf{S}^{-1} \nabla_{\theta_{j+1}} \mathscr{L}_{j+1},
\end{equation}
where $\eta$ is the learning rate and $\mathbf{S}$ is the Fisher information matrix (FIM). With the gradient matrix $O$ defined above, the empirical FIM is
\begin{equation}\label{eq:15}
\begin{aligned}
\mathbf{S} &= \mathbb{E}_{\mathbf{x} \sim \hat{P}_{j+1}^{\theta_{j+1}}} [\nabla_{\theta_{j+1}} \ln \hat{P}_{j+1}^{\theta_{j+1}}(\mathbf{x}) \nabla_{\theta_{j+1}} \ln \hat{P}_{j+1}^{\theta_{j+1}}(\mathbf{x})^\top] \\ &\approx O^{\top}O.
\end{aligned}
\end{equation}
Because $O^\top O$ can be singular when the number of parameters exceeds the batch size, we use the damped natural-gradient form
\begin{equation}\label{eq:16}
    \delta \theta_{j+1} = -\eta (O^{\top}O + \lambda_d \mathbf{I})^{-1}O^{\top}\boldsymbol{\mathcal{R}},
\end{equation}
where $\lambda_d>0$ is a damping parameter. The damping improves numerical stability and avoids requiring the unregularized FIM to be invertible. Applying the Woodbury identity gives the equivalent batch-space form
\begin{equation}\label{eq:17}
    \begin{aligned}
    \delta\theta_{j+1}
    &= -\frac{\eta}{\lambda_d}\left[O^{\top}\boldsymbol{\mathcal{R}} - O^\top (OO^\top + \lambda_d \mathbf{I})^{-1}OO^{\top}\boldsymbol{\mathcal{R}}\right].
    \end{aligned}
\end{equation}
This form avoids inversion of the $N_P \times N_P$ parameter-space matrix and instead requires solving a linear system in the $N_B \times N_B$ batch space.

We implement Eq.~\eqref{eq:17} by exact block accumulation, as illustrated schematically in Fig.~\ref{fig:main}(d). The batch is divided into $B$ sub-batches $\{\mathcal{B}_b\}_{b=1}^B$, and the corresponding gradient matrix blocks are $O^{(b)} \in \mathscr{R}^{N_b \times N_P}$, with $\sum_b N_b=N_B$. The full matrix $\mathrm{D}=OO^\top$ is not approximated by a block diagonal matrix; all cross-block products are included as

\begin{equation}\label{eq:18}
    \mathrm{D}_{bl}=O^{(b)}(O^{(l)})^\top, \qquad b,l=1,\ldots,B,
\end{equation}
where $\mathrm{D}_{lb}=\mathrm{D}_{bl}^\top$ is used to avoid redundant computation, reducing the number of block-matrix products by nearly $50\%$ for large $B$. Thus the assembled $\mathrm{D}$ is exactly the same matrix that would be obtained by first concatenating all blocks into $O$ and then forming $OO^\top$, while the peak memory is controlled by the chosen block sizes.

After constructing $\mathrm{D}$, we compute
\begin{equation}\label{eq:19}
    g=O^\top \boldsymbol{\mathcal{R}}=\sum_{b=1}^B (O^{(b)})^\top \boldsymbol{\mathcal{R}}^{(b)}, \qquad c=\mathrm{D}\boldsymbol{\mathcal{R}},
\end{equation}
and solve the damped batch-space linear system
\begin{equation}\label{eq:20}
    (\mathrm{D}+\lambda_d \mathbf{I})u=c.
\end{equation}
The final update is then evaluated block by block as
\begin{equation}\label{eq:21}
    \delta\theta_{j+1}=-\frac{\eta}{\lambda_d}\left[g-\sum_{b=1}^B (O^{(b)})^\top u^{(b)}\right],
\end{equation}
where $u^{(b)}$ is the segment of $u$ associated with sub-batch $b$. Because Eq.~\eqref{eq:18} includes every pair of sub-batches, this blocked procedure is algebraically equivalent to the full Woodbury update in Eq.~\eqref{eq:17}; blocking only changes the memory layout and computation schedule, not the natural-gradient direction. For larger systems, the same exact accumulation is performed in a streaming mode, in which each block of $O$ is computed, accumulated, and released before the next block is processed. An adaptive three-tier policy selects full caching, large-block streaming, or a small-block fallback according to the available GPU memory. Reducing the per-block memory limit lowers peak memory and extends the system sizes accessible on a single GPU, at the cost of additional blockwise operations. Fig.~\ref{fig:scaling} quantifies this memory-computation tradeoff and the resulting scaling with lattice size.

For the 2D finite-time calculations, blocking introduces a modest time overhead for MADE but reduces its memory-scaling exponent from $3.44$ to $3.07$ and avoids storing the full gradient matrix at once. The Transformer calculations likewise use blocked natural-gradient optimization and exhibit substantially weaker time and parameter scaling. With a 20GB memory limit for each gradient block and otherwise unchanged training settings, a single 80GB H100 reaches $L=10$ ($N=100$) with MADE and $L=9$ ($N=81$) with the Transformer. MADE is therefore practical for the smaller systems considered here, whereas the Transformer becomes favorable beyond approximately $N=100$~\cite{zhong2026scalable}. Accordingly, the 3D calculation at $L=5$ ($N=125$) uses the Transformer architecture detailed in Appendix~\ref{app:Computational cost}. Extrapolating the measured Transformer memory scaling, distributing the gradient blocks over four to eight 80GB GPUs is projected to extend the accessible range to approximately $L=15$-$19$ ($N\approx225$-$361$); a comparable site count in 3D corresponds to about $L=6$-$7$ ($N=216$-$343$). These estimates are resource projections rather than demonstrated benchmarks and may be reduced by communication and architecture-dependent overheads. The present size ceiling is therefore primarily a computational-resource limit of the current implementation, and can be pushed further through additional GPUs, smaller gradient blocks, or streaming accumulation without changing the exact blocked natural-gradient construction in Eqs.~\eqref{eq:18}-\eqref{eq:21}.

\section{Applications}\label{sec:Applications}
\begin{figure*}
\includegraphics[width=1\textwidth]{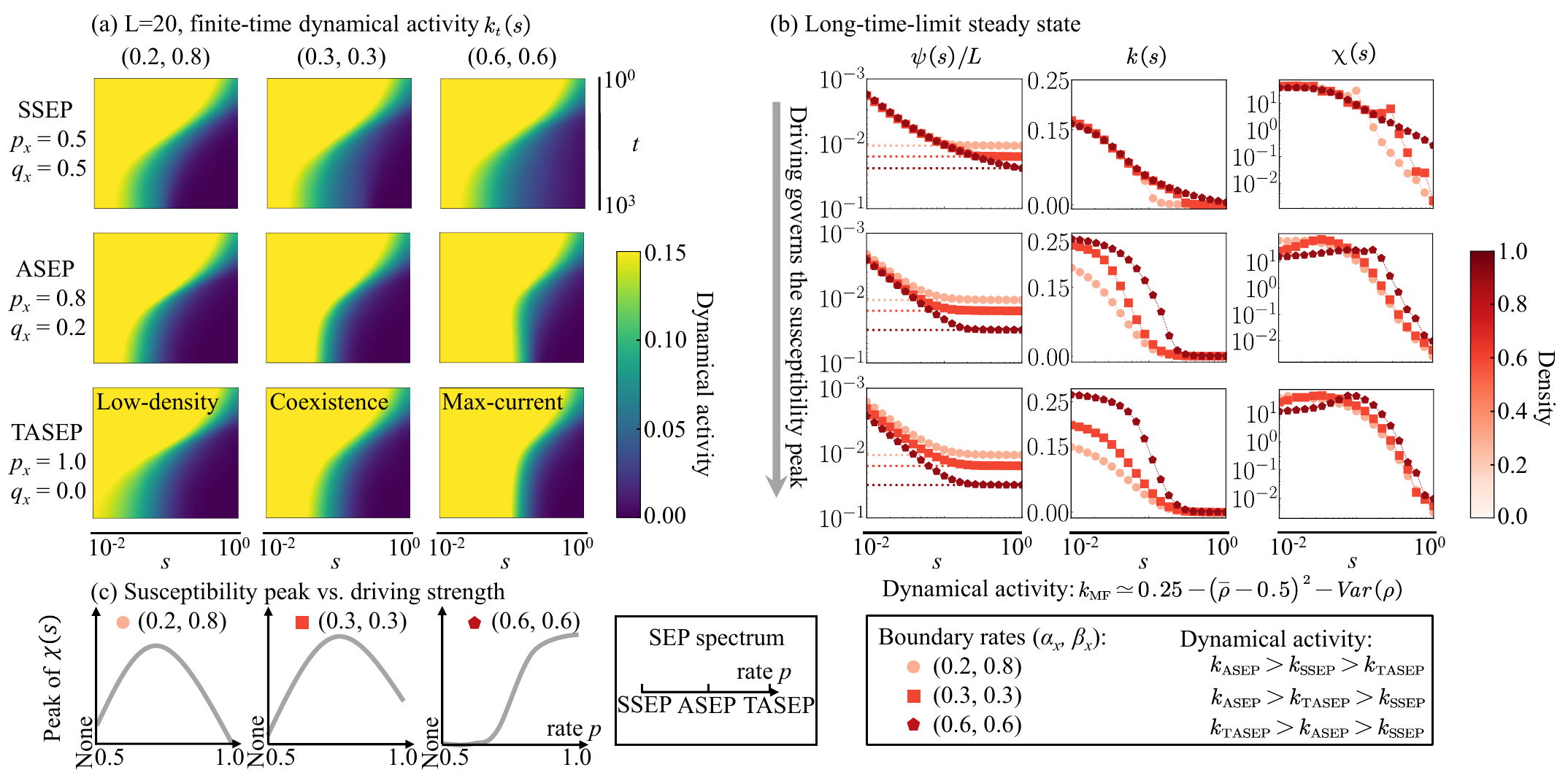}
\caption{\label{fig:1DASEP_boundary_effect_merge}\textbf{Finite-time activity and boundary and bulk effects in 1D SEPs.} (a) Finite-time evolution of dynamical activity $k_t(s)$ for $L = 20$, illustrating the crossover from an active phase (yellow) to an inactive phase (dark blue) across the counting field $s$ and trajectory time $t$. Rows display SSEP ($p_x = q_x = 0.5$), ASEP ($p_x = 0.8, q_x = 0.2$), and TASEP ($p_x = 1, q_x = 0$). Columns represent three transport regimes in 1D TASEP defined by boundary rates $(\alpha_x, \beta_x)$: low-density $(0.2, 0.8)$, coexistence $(0.3, 0.3)$, and maximal-current $(0.6, 0.6)$. The narrowing of dark inactive regions reflects the boundary-induced transition from low-density to maximal-current phases. (b) Steady-state SCGF $\psi(s)/L$, dynamical activity $k(s)$, and dynamical susceptibility $\chi(s)$ under varying bulk driving and boundary rates. In column $\chi(s)$, top-to-bottom rows show susceptibility curves of the three transport regimes under SSEP, ASEP, and TASEP. (c) Susceptibility peak heights as a function of bulk driving, illustrating how the peak heights vary in each regime. The bottom-right legend defines the markers and ranks the steady-state activity $k(s)$ under SSEP, ASEP, and TASEP at a weak counting field $s = 10^{-2}$, reflecting the consistent ranking of dynamical activity obtained from both the mean-field approximation $k_{\mathrm{MF}}$ and VAN.
}
\end{figure*}
This section organizes the physical results around dynamical activity and two related connections: the density-activity relation and the activity susceptibility response. The 1D results establish the reference correspondence between density phases and activity, after which we focus on the directional-density organization in 2D and the finite-time active-inactive scaling in 2D and 3D. The directional-density criterion and its numerical support are presented in Appendix~\ref{app:2d-directional-convergence} and Table~\ref{tab:2dtasep-numerical}, while the corresponding benchmarks are documented in Appendices~\ref{app:1D SSEP benchmark}, \ref{app:2d-ssep-benchmark}, \ref{app:3d-ssep-vmc}, and~\ref{app:finite-time-partition-validation}.

\subsection{1D: boundary and bulk effects}\label{subsec:Boundary effect on transport efficiency and jamming in one-dimensional system}

\begin{figure*}
\includegraphics[width=1\textwidth]{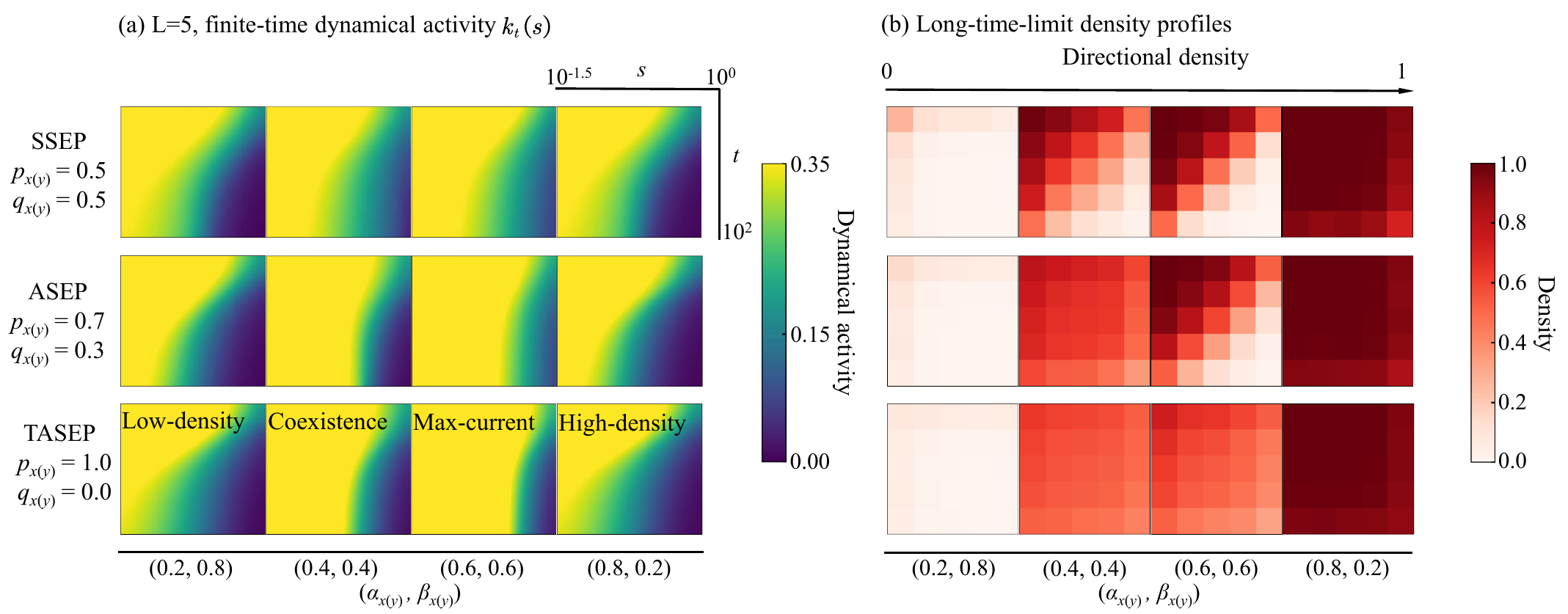}
\caption{\label{fig:dynamical_activity_density_profiles}\textbf{Crossover from diffusive to ballistic regimes in 2D SEPs and spatial density profiles.} (a) Finite-time dynamical activity $k_t(s)$ as a function of the counting field $s$ and trajectory time $t$ for system size $L = 5$. Rows represent symmetric (SSEP, $p_{x(y)} = q_{x(y)} = 0.5$), asymmetric (ASEP, $p_{x(y)} = 0.7, q_{x(y)} = 0.3$), and totally asymmetric (TASEP, $p_{x(y)} = 1.0, q_{x(y)} = 0.0$) bulk driving. From left to right, as the directional density gradually increases across the columns, the active phase (light-yellow) first broadens and then narrows, whereas the inactive phase (dark-blue) correspondingly narrows and then widens. Columns represent four typical boundary rate pairs $(\alpha_{x(y)}, \beta_{x(y)})$. (b) Corresponding density profiles ($t=50, s=10^0$). Despite sharing comparable dynamical activity, the density profiles reveal distinct microscopic mechanisms underlying distinct inactive phases: extraction-dominated boundaries, e.g., $(0.2, 0.8)$, induce low-density regions (white), whereas injection-dominated boundaries, e.g., $(0.8, 0.2)$, trigger macroscopic high-density regions (dark-red). Under a stronger bulk drive, the particle density modulates along the driving direction, thereby shaping distinct density profiles.}
\end{figure*}

We first determine how boundary rates and bulk driving separately govern the dynamical response in one dimension. In the matrix-product ansatz of Derrida et al.~\cite{31_Derrida1992,31_derrida1993exact}, the open TASEP is classified into low-density (LD), high-density (HD), and maximal-current (MC) phases, together with a coexistence line in Fig.~\ref{fig:1DTASEP_phase_graph}, according to the boundary-rate pair $(\alpha_x,\beta_x)$. Steady-state current-activity fluctuations and their dependence on these boundary-induced phases have been analyzed previously~\cite{112_stinchcombe2012statistics,113_dequeiroz2012current}. For the dynamical observables considered here, the LD and HD phases exhibit similar phase boundaries, with both showing relatively low activity compared with the MC phase. Similar behavior is observed in the 2D systems shown in Fig.~\ref{fig:dynamical_activity_density_profiles}(a). Their microscopic mechanisms nevertheless differ: inactive behavior in the LD phase originates from particle scarcity, whereas that in the HD phase stems from particle jamming, as illustrated by the analogous density profiles in Fig.~\ref{fig:dynamical_activity_density_profiles}(b). Accordingly, Fig.~\ref{fig:1DASEP_boundary_effect_merge}(a) uses the LD phase as a representative low-activity case and does not show the HD case, which exhibits similar activity behavior.

The three columns of Fig.~\ref{fig:1DASEP_boundary_effect_merge}(a) use representative TASEP boundary-rate pairs from the LD, coexistence, and MC regimes: $(\alpha_x,\beta_x)=(0.2,0.8)$, $(0.3,0.3)$, and $(0.6,0.6)$, respectively. For each fixed boundary pair, we vary the bulk hopping rates across the rows, obtaining the SSEP ($p_x=q_x=0.5$) in the first row, the ASEP ($p_x=0.8$, $q_x=0.2$) in the second row, and the TASEP ($p_x=1.0$, $q_x=0.0$) in the third row. These choices yield finite-time dynamical-activity maps across both boundary and bulk rates. In the TASEP row, moving from left to right through the LD, coexistence, and MC regimes progressively expands the yellow active region and narrows the dark-blue inactive region, establishing a direct correspondence between the finite-time activity landscape and the steady-state density phase.

The long-time observables in Fig.~\ref{fig:1DASEP_boundary_effect_merge}(b) show, within each row, the SCGF $\psi(s)/L$, activity $k(s)$, and susceptibility $\chi(s)$ for a fixed bulk hopping rate with different boundary conditions. In the diffusive SSEP, the low-activity boundary rates $(\alpha_x,\beta_x)=(0.2,0.8)$ (circles) produce the pronounced susceptibility peak. In the ballistic TASEP, by contrast, the high-activity boundary rates $(0.6,0.6)$ (pentagons) produce the dominant peak. Meanwhile, Comparing the rows shows that increasing the bulk driving modulates the susceptibility response across the different boundary regimes. Fig.~\ref{fig:1DASEP_boundary_effect_merge}(c) makes this trend explicit: as the right hopping rate increases from $p_x=0.5$ to $1.0$, the peak associated with the low-activity boundary rates (circles) first increases, then decreases, and eventually disappears, whereas the peak associated with the high-activity boundary rates (pentagons) increases progressively.

The bottom-right panel of Fig.~\ref{fig:1DASEP_boundary_effect_merge} provides a qualitative mean-field interpretation of the activity through $k_{\mathrm{MF}}\simeq0.25-(\bar\rho-0.5)^2-\mathrm{Var}(\rho)$; its derivation and approximations are given in Appendix~\ref{app:mean-field-activity}. For each fixed boundary pair, the activity ordering among the SSEP, ASEP, and TASEP predicted by this expression agrees with the VAN ordering at $t=10^3$ and $s=10^{-2}$. The expression shows that activity is enhanced when the mean density is closer to half filling and the density variance is smaller. Thus, a more homogeneous density profile near $\bar\rho=0.5$ generally supports higher activity, providing a qualitative explanation for how changes in the density fields influence the susceptibility response.

Together, these results show that the steady-state phase of the TASEP is reflected in its finite-time activity phase and reveal how the boundary rates and bulk driving separately govern the susceptibility response. The mean-field activity relation further explains the observed activity trends. We next examine how these boundary and bulk effects extend to coupled transport in 2D.

\begin{figure*}
\includegraphics[width=1\textwidth]{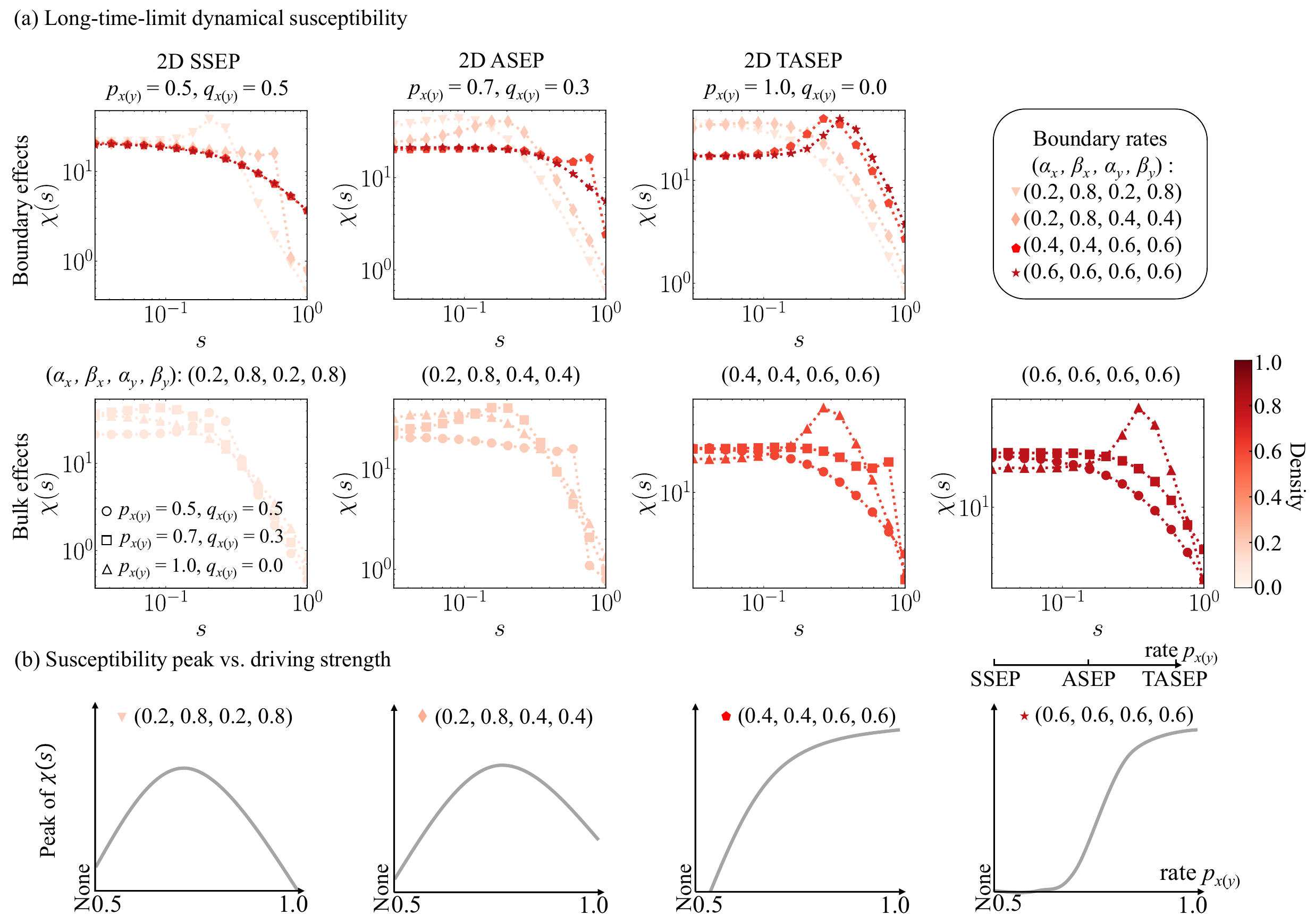}
\caption{\label{fig:2DASEP_current_density_steady}\textbf{Boundary and bulk effects in 2D SEPs.} (a) Top row: Steady-state susceptibility $\chi(s)$ in the long-time limit for $L = 5$ in the SSEP ($p_{x(y)} = q_{x(y)} = 0.5$), ASEP ($p_{x(y)} = 0.7, q_{x(y)} = 0.3$), and TASEP ($p_{x(y)} = 1.0, q_{x(y)} = 0.0$). Symbols represent four boundary combinations $(\alpha_x, \beta_x, \alpha_y, \beta_y)$ spanning low-activity (light) to high-activity (dark) regimes. In the diffusive SSEP, low-activity regimes exhibit stronger susceptibility responses, whereas increasing the bulk driving shifts the dominant response toward high-activity regimes in the ballistic TASEP limit. Bottom row: Dependence of the susceptibility response on bulk driving for the same four boundary combinations, progressing from SSEP (circles) through ASEP (squares) to TASEP (triangles). (b) Susceptibility peak heights as a function of bulk driving, illustrating how the peak heights vary in each regime. Different combinations of boundary and bulk rates reveal signatures of incipient dynamical phase transitions.
}
\end{figure*}
\subsection{2D: boundary and bulk effects}\label{subsec:Transverse-longitudinal flow coupling in two-dimensional system}

As illustrated by the 2D square lattice in Fig.~\ref{fig:main}(a), particles can hop along both the $x$ and $y$ directions. The coupling between transverse and longitudinal transport makes the problem more complex. To connect the 2D bulk density to the 1D boundary phases, we introduce a directional-density criterion (Appendix~\ref{app:2d-directional-convergence}): when the boundary conditions in both directions select the same steady-state density according to the 1D TASEP phase diagram, the 2D bulk density converges to this shared value in the thermodynamic limit. This criterion yields a three-phase organization analogous to that in one dimension. As shown schematically in Fig.~\ref{fig:main}(e-1), the low-density, maximal-current, and high-density phases appear successively along the diagonal defined by the two directional densities.
\begin{table}[t]
\centering 
\caption{\label{tab:2dtasep-numerical}VAN numerical results for the density ($\hat\rho$) and directional currents ($\hat J_x,\hat J_y$) of the 2D TASEP at $s=10^{-3}$ and $t=50$. The lattice size is $L=5$. Values in parentheses denote the bulk-density and current predictions inherited from the 1D boundary-selected phases. For coexistence, $\bar\rho$ denotes the spatially averaged density.}
\footnotesize
\begin{tabular}{ccc}
\toprule
\makecell[c]{Phase \\ type} & \makecell[c]{Density \\ $\hat{\rho}$ ($\rho$)} & \makecell[c]{Currents \\ $\hat{J}_x, \hat{J}_y$ ($J_x, J_y$)} \\
\midrule
\makecell[c]{Low-density\\$\alpha_{x(y)}=0.1$, \\ $\alpha_{x(y)}<\beta_{x(y)}$} &
\makecell[c]{$\hat\rho=0.105$ \\ $  (\rho=0.10)$} &
\makecell[c]{$\hat{J}_x =0.091,\hat{J}_y=0.093$ \\ ($J_x=J_y=0.09$)}  \\
\addlinespace[0.5em]

\makecell[c]{Coexistence \\ $\alpha_{x} = \beta_{x}=0.4$ \\ $\alpha_{y} = \beta_{y}=0.4$} &
\makecell[c]{$\bar\rho=0.500$ \\ $  (\bar\rho=0.50)$} &
\makecell[c]{$\hat{J}_x =0.231,\hat{J}_y=0.229$ \\ ($J_x=J_y=0.24$)}  \\
\addlinespace[0.5em]

\makecell[c]{Maximal-current \\ $\alpha_{x},\beta_{x} \ge 0.5$ \\ $\alpha_{y},\beta_{y} \ge 0.5$} &
\makecell[c]{$\hat\rho=0.481$ \\ $  (\rho=0.50)$} &
\makecell[c]{$\hat{J}_x =0.263,\hat{J}_y=0.284$ \\ ($J_x=J_y=0.25$)} \\
\addlinespace[0.5em]
\makecell[c]{High-density\\$\beta_{x(y)}=0.3$ \\$\alpha_{x(y)}>\beta_{x(y)}$} &
\makecell[c]{$\hat\rho=0.654$ \\ $  (\rho=0.70)$} &
\makecell[c]{$\hat{J}_x =0.208,\hat{J}_y=0.217$ \\ ($J_x=J_y=0.21$)} \\
\bottomrule
\end{tabular}
\end{table}

Based on this criterion, we select one representative boundary-rate set $(\alpha_x,\beta_x,\alpha_y,\beta_y)$ from each phase: $(0.1,0.4,0.1,0.9)$ for the low-density phase, $(0.4,0.4,0.4,0.4)$ for coexistence, $(0.6,0.7,0.8,0.9)$ for the maximal-current phase, and $(0.4,0.3,0.6,0.3)$ for the high-density phase. For these four boundary conditions, we first use a mean-field approximation to calculate the steady-state density and directional currents of the 2D TASEP (Appendix~\ref{app:2d-directional-convergence}, Table~\ref{tab:2dtasep-numerical-MFA}). The VAN calculation then evolves the full 2D dynamics and directly samples the corresponding biased ensembles at $s=10^{-3}$ and $t=50$. In Table~\ref{tab:2dtasep-numerical}, the VAN estimates for the four representative phases are close to the mean-field values, supporting the directional-density criterion; the detailed comparison and the $L=30$ mean-field results are given in Appendix~\ref{app:2d-directional-convergence}. Guided by this criterion, Fig.~\ref{fig:dynamical_activity_density_profiles}(a) orders the boundary conditions by increasing directional density from left to right. The bottom TASEP row displays the corresponding finite-time activity landscape, while the middle and top rows show the ASEP and SSEP results under the same boundary conditions. The low- and high-density phases exhibit similar low-activity behavior, but the density profiles in Fig.~\ref{fig:dynamical_activity_density_profiles}(b) reveal distinct microscopic mechanisms in the inactive phase: particle scarcity in the low-density phase and jamming in the high-density phase.
\begin{figure*}
\includegraphics[width=1\textwidth]{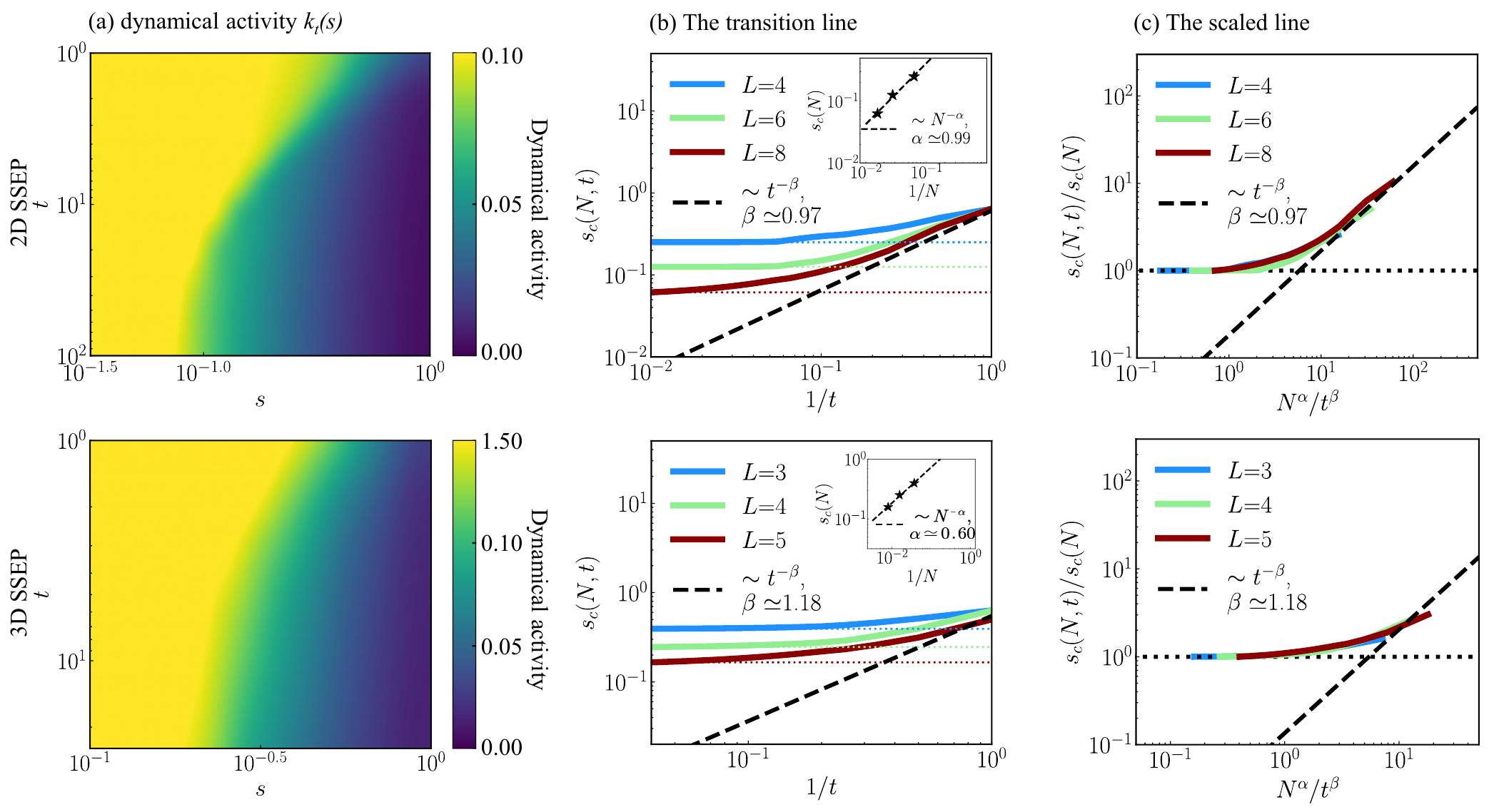}
\caption{\label{fig:2DSSEP_dynamical_activity}\textbf{The dynamical active-inactive phase transition of 2D and 3D SSEP.} The top and bottom rows present results for 2D SSEP on square lattices and 3D SSEP on cubic lattices, respectively. (a) The dynamical activity $k_t(s)$ reveals the phase transition versus trajectory time $t$ and the counting field $s$, with $L = 8$ for 2D and $L = 5$ for 3D, illustrating the crossover from an active phase (light-yellow) to an inactive phase (dark-blue). (b) The critical point $s_c(N,t)$ over time, giving critical exponents $\alpha$ from the steady-state scaling $s_c(N) \sim N^{-\alpha}$ (shown in the insets and the horizontal dotted lines) and $\beta$ from the finite-time power-law scaling $s_c(N,t) - s_c(N) \sim t^{-\beta}$ (the black dashed lines). The extracted exponents are $\alpha \approx 0.99$, $\beta \approx 0.97$ for 2D, and $\alpha \approx 0.60$, $\beta \approx 1.18$ for 3D. (c) The scaled transition lines, with $s_c(N,t)/s_c(N)$ plotted against $N^{\alpha}/t^{\beta}$, show a consistent collapse that supports a common finite-time scaling form over the accessible system sizes. Parameters: a unit bulk hopping rate and boundary rates of $0.5$, with $L \in \{4,6,8\}$ for 2D SSEP ($N = L^2$) and $L \in \{3,4,5\}$ for 3D SSEP ($N = L^3$).
}
\end{figure*}

In the bottom TASEP row of Fig.~\ref{fig:dynamical_activity_density_profiles}(a), moving from the low-density phase toward the maximal-current phase, the increasing density is accompanied by a progressive expansion of the active region, mirroring the behavior of the 1D TASEP in Fig.~\ref{fig:1DASEP_boundary_effect_merge}(a). This reveals a correspondence between the density phases and the active-inactive dynamical phase. In the third column of Fig.~\ref{fig:dynamical_activity_density_profiles}(a), as the bulk dynamics changes from diffusive SSEP through driven ASEP to ballistic TASEP, the active region in the finite-time activity map expands progressively, showing that stronger bulk drive enhances activity under maximal-current boundary conditions (Fig.~\ref{fig:main}(e-2)). The corresponding density profiles in Fig.~\ref{fig:dynamical_activity_density_profiles}(b) change from a density-gradient structure in the diffusive system toward a smoother and more uniform bulk profile near half filling as the drive increases. These profiles are evaluated at $s=10^{0}$ and $t=50$; the transient snapshots in Appendix~\ref{app:2d-ssep-benchmark} and Fig.~\ref{fig:2_2DASEP_spatial_structure} show that the spatial structures are already nearly unchanged by $t=12.5$. Thus, the boundary rates select the density phase, whereas the bulk driving modulates the spatial organization.

The top row of Fig.~\ref{fig:2DASEP_current_density_steady}(a) shows the long-time dynamical susceptibility $\chi(s)$ for a fixed bulk hopping rate and different boundary conditions. In the diffusive SSEP, the low-activity boundary set $(\alpha_x,\beta_x,\alpha_y,\beta_y)=(0.2,0.8,0.2,0.8)$, denoted by inverted triangles, exhibits a pronounced susceptibility response. In the ballistic TASEP, by contrast, the high-activity boundary set $(0.6,0.6,0.6,0.6)$, denoted by stars, exhibits a pronounced susceptibility peak. The corresponding activity curves are shown in Appendix~\ref{app:2d-ssep-benchmark}: the top row of Fig.~\ref{fig:2DSEP_activity_long-time} compares boundary conditions at fixed bulk dynamics, while the bottom row compares bulk driving at fixed boundary conditions. Comparing the three panels therefore shows that increasing the bulk drive shifts the dominant susceptibility response from low-activity to high-activity boundary conditions.

The bottom row of Fig.~\ref{fig:2DASEP_current_density_steady}(a) presents the same susceptibility curves for each fixed boundary-rate set as the bulk dynamics changes from the SSEP (circles) through the ASEP (squares) to the TASEP (triangles). Fig.~\ref{fig:2DASEP_current_density_steady}(b) makes the corresponding peak-height trends explicit. For the two lighter, low-activity boundary sets, the susceptibility peak first increases and then decreases as the bulk drive strengthens. For the two darker, high-activity sets, the peak increases progressively toward the TASEP limit. Thus, the boundary and bulk effects separately govern the dynamical susceptibility response.

\subsection{2D: dynamical phase transitions}\label{subsec:Dynamical phase transitions in two-dimensional system}

We next examine the active-inactive dynamical phase transition in the 2D SSEP on an open square lattice. The bulk hopping rates are $p_x=q_x=p_y=q_y=1.0$, and all injection and extraction rates are fixed at $0.5$. For a lattice of linear size $L$, the configuration space contains $2^N$ states with $N=L^2$, motivating the use of the VAN for the finite-time calculations. The VAN accurately captures the phase transition. The calculated dynamical partition function coincides with the numerically exact values available for the small system sizes (Appendix~\ref{app:finite-time-partition-validation}, Fig.~\ref{fig:tracking-time-evolution-dynamical-partition-function}). Our result at long time matches the steady-state estimate from the variational Monte Carlo (VMC) method (Appendix~\ref{app:2d-ssep-benchmark}, Fig.~\ref{fig:comparison result2_1s}). For the finite-time regime, which was previously explored only for 1D~\cite{12_causer2022finite}, our result (Appendix~\ref{app:1D SSEP benchmark}, Fig.~\ref{fig:1DSSEP}) agrees with tensor networks. For the 2D SSEP at finite time, we obtain the dynamical activity $k_t(s)$ as a function of time $t$ and the counting field $s$ with $s > 0$, showing two phases with extensive or subextensive activities (top row of Fig.~\ref{fig:2DSSEP_dynamical_activity}(a)).

The critical point $s_c(N,t)$ as a function of system size $N$ and time $t$ can be identified numerically by the peak position of the dynamical susceptibility $\chi_t(s)$, as summarized schematically in Fig.~\ref{fig:main}(e-3). Here and below, the unsubscripted $\alpha$ and $\beta$ denote the finite-size and temporal scaling exponents, whereas $\alpha_\mu$ and $\beta_\mu$ denote boundary rates along direction $\mu$. We conduct a scaling analysis of the critical point $s_c(N,t)$ for the 2D SSEP shown in the top row of Fig.~\ref{fig:2DSSEP_dynamical_activity}. In the long-time regime, the scaling of system size $s_c(N) \sim N^{-\alpha}$ gives the exponent $\alpha \approx 0.99$ for 2D [inset of Fig.~\ref{fig:2DSSEP_dynamical_activity}(b)]. By inspecting the short-time regime, the critical point approximately scales as $s_c(t) \sim t^{-1}$. These two regimes motivate the approximation $s_c(N,t) \approx s_c(N) + s_c(t)$. We then estimate the temporal scaling by fitting $\ln[s_c(N,t) - s_c(N)]$ versus $-\ln(t)$, giving $s_c(t) \sim t^{-\beta}$ with $\beta \approx 0.97$ for 2D (Fig.~\ref{fig:2DSSEP_dynamical_activity}(b)), which is close to unity. The 2D finite-size exponent is also consistent with the approach to the steady-state transition reported in~\cite{19_causer2023optimal} over the studied sizes. Dividing $s_c(N,t)$ by $s_c(N)$ and rescaling time as $N^{\alpha}t^{-\beta}$ produces the collapse shown in Fig.~\ref{fig:2DSSEP_dynamical_activity}(c).

\begin{table*}[ht!]
\caption{\label{tab:contributions}Summary of validation, physical findings, and transport mechanisms in each dimension.}
\footnotesize
\setlength{\tabcolsep}{0pt}
\renewcommand{\arraystretch}{1.25}
\begin{tabular*}{\textwidth}{@{\extracolsep{\fill}}c l l l@{}}
\toprule
\multicolumn{1}{c}{\makecell[c]{System}} & \multicolumn{1}{c}{\makecell[c]{Benchmarks}} & \multicolumn{1}{c}{\makecell[c]{Main physical finding}} & \multicolumn{1}{c}{\makecell[c]{Transport mechanism}} \\
\midrule
\makecell[c]{1D SEPs} &
\begin{tabular}[c]{@{}l@{}}(1) Exact MPA density and current,\\ \phantom{(2) }relative errors within $1.2\%$;\\ (2) 1D SSEP finite-time MPS~\cite{12_causer2022finite}.\end{tabular} &
\begin{tabular}[c]{@{}l@{}}Density-activity phase correspondence;\\ boundary and bulk effects;\\ finite-time VAN reaches $L=40$.\end{tabular} &

\begin{tabular}[c]{@{}l@{}}Low activity are more susceptible in\\ diffusive dynamics; high activity are \\ more susceptible in ballistic dynamics.\end{tabular} \\
\addlinespace[0.5em]
\makecell[c]{2D SEPs} &
\begin{tabular}[c]{@{}l@{}}(1) Exact $L=3$ finite-time evolution,\\ \phantom{(2) }errors $<10^{-4}$ in tested cases;\\ (2) tensor-network steady states~\cite{19_causer2023optimal}. \end{tabular} &
\begin{tabular}[c]{@{}l@{}}Directional-density criterion relates 1D \\ TASEP to 2D TASEP density phases;\\ VAN from $N\leq36$~\cite{17_tang2024learning} to $N\leq100$.\end{tabular} &
\begin{tabular}[c]{@{}l@{}}In the maximal-current sector,\\ stronger bulk drive enhances activity\\ and reshapes spatial density structure.\end{tabular} \\
\addlinespace[0.5em]
\makecell[c]{3D SSEP} &
\begin{tabular}[c]{@{}l@{}}(1) Exact $L=2$ finite-time evolution,\\ \phantom{(2) }errors $<10^{-5}$ in tested cases;\\ (2) steady-state VMC comparison.\end{tabular} &
\begin{tabular}[c]{@{}l@{}}Finite-size and finite-time scaling\\ of $s_c(N,t)$ up to $L=5$ ($N=125$);\\ collapse with $\beta\gtrsim1$, $\alpha\approx0.60$.\end{tabular} &
\begin{tabular}[c]{@{}l@{}}Critical-field scaling is broadly consistent\\ with diffusive length-scale control,\\ $s_c\sim L^{-2}$, with finite-size corrections.\end{tabular} \\
\bottomrule
\end{tabular*}
\end{table*}
\subsection{3D: dynamical phase transitions}\label{subsection:Dynamical activity and phase boundaries in three-dimensional system}

We next ask whether the finite-time scaling picture established for the 2D SSEP persists in a genuinely 3D geometry. To this end, we study the 3D SSEP on cubic lattices with open boundaries. As in the 2D SSEP dynamical-transition analysis, we use unit bulk hopping rates in each direction and fix all boundary rates at $0.5$. The configuration space now grows as $2^N$ with $N=L^3$, making the calculation substantially more demanding than in two dimensions. Nevertheless, the VAN approach resolves the finite-time activity landscape for the accessible 3D systems. Its dynamical partition function agrees with numerically exact results for $L=2$ (Appendix~\ref{app:finite-time-partition-validation}, Fig.~\ref{fig:3d-validation-exact-solutions}), and its long-time finite-size trend is consistent with the steady-state variational Monte Carlo (VMC)~\cite{15_carleo2017solving,17_tang2024learning} estimate, which gives a comparable exponent ($\alpha \approx 0.61$) (Appendix~\ref{app:3d-ssep-vmc}, Fig.~\ref{fig:comparison-3dresult-1s}). As shown in the bottom row of Fig.~\ref{fig:2DSSEP_dynamical_activity}(a), the dynamical activity $k_t(s)$ again separates active and inactive trajectory regimes as the counting field $s>0$ is varied.

The critical point $s_c(N,t)$ as a function of system size and time can be identified numerically by the peak position of the dynamical susceptibility $\chi_t(s)$. We conduct a scaling analysis of the critical point $s_c(N,t)$ for the 3D data shown in the bottom row of Fig.~\ref{fig:2DSSEP_dynamical_activity}. In the long-time regime, the scaling of system size $s_c(N) \sim N^{-\alpha}$ gives the exponent $\alpha \approx 0.60$ for 3D [inset of Fig.~\ref{fig:2DSSEP_dynamical_activity}(b)]. By inspecting the finite-time regime, the critical counting field approaches its steady-state value according to $s_c(N,t)-s_c(N) \sim t^{-\beta}$. We estimate the temporal scaling by fitting $\ln[s_c(N,t)-s_c(N)]$ versus $-\ln(t)$, giving $\beta \gtrsim 1$ for 3D (Fig.~\ref{fig:2DSSEP_dynamical_activity}(b)). By dividing $s_c(N,t)$ by $s_c(N)$ and rescaling time as $N^{\alpha}/t^{\beta}$, the 3D $s_c(N,t)$ curves collapse together (Fig.~\ref{fig:2DSSEP_dynamical_activity}(c)), supporting the finite-time scaling form within the simulated ranges.

Combining the finite-size estimates suggests a systematic dimensional trend. We obtain $\alpha \approx 1.97$ in 1D (Appendix~\ref{app:1D SSEP benchmark}, Fig.~\ref{fig:1DSSEP}(b)), $\alpha \approx 0.99$ in 2D (top row of Fig.~\ref{fig:2DSSEP_dynamical_activity}(b)), and a steady-state 3D VMC estimate $\alpha \approx 0.61$ (Appendix~\ref{app:3d-ssep-vmc}, Fig.~\ref{fig:comparison-3dresult-1s}(b)). The finite-time 3D fit gives $\alpha \approx 0.60$ over $L\in\{3,4,5\}$ (bottom row of Fig.~\ref{fig:2DSSEP_dynamical_activity}(b)), indicating stronger finite-time and finite-size corrections. The 1D and 2D estimates are consistent with available tensor-network benchmarks, which reported $\alpha \approx 2.01$ in 1D~\cite{12_causer2022finite} and a comparable steady-state 2D exponent around $\alpha \approx 0.92$~\cite{19_causer2023optimal}.

Because $N=L^d$, the dimensional trend above suggests the approximate relation
\begin{equation}
s_c\sim N^{-2/d}=L^{-2}, \qquad \alpha=\frac{2}{d}.
\label{eq:mft-dimensional-scaling}
\end{equation}
Our numerical estimates provide support for this relation between the critical counting field and the characteristic diffusive length scale, consistent with the universal finite-size scaling of activity fluctuations reported in~\cite{86_appert2008universal} and with analytical studies of the activity-biased SSEP~\cite{81_lecomte2012inactive}.

\section{\label{sec:Discussion}Discussion}

We have presented a variational framework that tracks the finite-time tilted dynamics of simple exclusion processes and evaluates the dynamical partition function without enumerating the full configuration space. The framework combines an autoregressive probability representation with blocked natural-gradient descent, replacing the Adam optimization used in the earlier finite-time VAN formulation~\cite{17_tang2024learning}; controlled natural-gradient VAN benchmarks have shown faster convergence and more accurate variational estimates than conventional Adam optimization~\cite{20_Liu2025Efficient}. The physical results follow a common hierarchy: boundary and bulk rates determine the density structure, the density structure organizes the dynamical activity, and the susceptibility reveals where the activity becomes strongly responsive and a dynamical transition may emerge. In 1D TASEP, representative low-density, coexistence, and maximal-current boundary regimes produce corresponding structures in the activity maps, while in 2D TASEP the directional-density criterion connects the low-density, high-density, and maximal-current densities selected by the 1D phase diagram to matched two-dimensional sectors. For the SSEP, susceptibility peaks locate the active-inactive crossover, and the explicit finite-time evolution in 2D and 3D reveals new finite-time and finite-size scaling relations beyond one dimension. Across 1D, 2D, and 3D, the resulting estimates are consistent with $s_c\sim L^{-2}$ over the accessible sizes, although larger 3D systems are needed to quantify the remaining finite-size corrections.

The accuracy of the finite-time evolution is controlled by the short-time operator and the variational projection at each time slice. We use a fixed MADE network of depth $4$ and width $16$, with $N_B=1000$, $\eta=1$, and $\lambda_d=10^{-5}$; the time step is $\delta t=0.05$ in 1D and $0.005$ in 2D and 3D, as summarized in Table~\ref{tab:van_fullwidth}. This MADE architecture is used for the finite-time calculations studied here with up to $N=64$ lattice sites. For larger systems, we replace MADE with a Transformer-based autoregressive representation~\cite{zhong2026scalable}, which offers greater expressiveness and slower memory growth at the cost of increased computation per epoch; we retain the natural-gradient optimizer rather than the Muon optimizer employed therein. Calculations beyond the tested range may require greater network depth or width to retain sufficient representational capacity. These architecture choices determine the accessible system size, whereas temporal accuracy depends on how the variational state is propagated between times.

In the present time-slice scheme, the first slice is trained for about $500$ epochs, whereas each subsequent slice is optimized for $3$ epochs. With this training schedule, the relative error of the dynamical partition function remains at the $10^{-4}$ level throughout the tested 2D evolution (Fig.~\ref{fig:tracking-time-evolution-dynamical-partition-function}) and below $10^{-5}$ in the 3D $L=2$ benchmark against exact numerical evolution (Fig.~\ref{fig:3d-validation-exact-solutions}). The agreement between the extrapolated finite-time SCGF and the independently optimized VMC values in Fig.~\ref{fig:3d-validation-exact-solutions} supports the effectiveness of the method, indicating that the finite-time estimates can be consistently extrapolated toward the steady state. Related time-dependent neural approaches include TDVP-based parameter evolution~\cite{91_reh2021time}, explicitly time-dependent ansatzes trained over an entire interval~\cite{vandewalle2025manybody}, and global-in-time Galerkin optimization~\cite{sinibaldi2026neural}. Compared with these interval-wide approaches, time-slice optimization provides direct local control and adapts naturally to transient changes, whereas global-time parameterizations share information across the trajectory and may reduce repeated optimization but require a more global temporal ansatz. The two strategies are complementary and could be combined through global-time propagation followed by time-slice natural-gradient refinement.
\begin{table*}[htbp]
\centering
\caption{Principal notation and conventions used throughout the paper. Subscripted $\alpha_\mu$ and $\beta_\mu$ denote boundary rates, whereas unsubscripted $\alpha$ and $\beta$ denote scaling exponents.}
\label{tab:notation}
\footnotesize
\begin{tabular*}{\textwidth}{@{\extracolsep{\fill}}clcl@{}}
\toprule
Symbol & Meaning & Symbol & Meaning \\
\midrule
$d,L,N=L^d$ & Dimension, linear size, and number of sites & $\mu\in\{x,y,z\}$ & Lattice direction \\
$p_\mu,q_\mu$ & Forward, backward hopping rates & $\alpha_\mu,\delta_\mu,\gamma_\mu,\beta_\mu$ & Boundary rates \\
$\mathbf{x}$ & Configuration vector & $x_i$ & Configuration variable at site $i$ \\
$a_i,a_i^\dagger$ & Annihilation and creation operators & $n_i$, $v_i$ & Particle-number and vacancy operators \\
$\mathbb{W},\mathbb{K},\mathbb{R}$ & Generator, transition, and escape matrices & $K,s,Z_t(s)$ & Activity, counting field, and partition function \\
$|P_t\rangle,|P_0\rangle,\omega_t$ & Probability vector, initial state, and trajectory & $\psi_t,k_t,\chi_t$ & SCGF, activity per site, and susceptibility \\
$\hat P_j^{\theta_j},\mathbb{T}_s,\mathscr{L}_j$ & VAN distribution, operator, and loss function & $N_{\mathrm{step}},N_B,N_P$ & number of time steps, batch size, network parameters \\
$O,\boldsymbol{\mathcal{R}},\mathbf{S},\mathrm{D}$ & gradient, reward, Fisher, and batch-space matrices & $\eta,\lambda_d$ & Learning rate and damping \\
$s_c(N,t),\alpha,\beta$ & Transition point and scaling exponents & $\rho,J,k_{\mathrm{MF}}$ & Mean density, current, and mean-field activity \\
\bottomrule
\end{tabular*}
\end{table*}

One computational challenge is the growing memory cost of natural-gradient optimization as the lattice size increases. The number of network parameters and intermediate activations grows rapidly with the number of lattice sites, making the representation of larger 2D and 3D probability distributions more demanding. Although the dimension of the batch-space matrix $\mathrm{D}=OO^\top$ is fixed by the batch size, constructing it exactly requires the gradient matrix $O$ with one column for every network parameter; storing its blocks and intermediate products can therefore use up the memory of a single GPU. This difficulty is compounded by the need to repeat the finite-time evolution over many values of $s$ (Table~\ref{tab:van_fullwidth}). These costs can be alleviated by distributing the gradient-matrix blocks and the construction of $\mathrm{D}$ across multiple GPUs, thereby reducing the peak memory on each device. The Transformer scaling projects an accessible range of $L\approx15$-$19$ ($N\approx225$-$361$),
comparable in site count to the $N=200$ systems studied using 1D finite-time matrix-product-state evolution~\cite{12_causer2022finite}, indicating that the present framework can reach system sizes representative of current finite-time studies (Sec.~\ref{subsection:Blocked natural-gradient descent}). Further efficiency gains may be achieved using time-dependent variational principles that evolve the network parameters directly and reduce the repeated optimization required at individual time slices~\cite{91_reh2021time,98_li2026weightflow}.

A possible extension is to finite-time current statistics across dimensions. Building on tensor-network studies of steady-state current and jammed-to-flowing transitions in the 2D ASEP~\cite{18_helms2020dynamical}, directional counting fields would allow the VAN evolution to track the finite-time SCGF, current, and susceptibility in each lattice direction for one- to three-dimensional SEPs. The resulting tilted distributions can also facilitate efficient sampling of rare trajectories~\cite{69_causer2021optimal,PhysRevE.111.034120}. Beyond SEPs, the framework may extend to reaction networks~\cite{tang2023neural,93_fang2023divide,92_Chuanbo_2024Distilling,weng2025tracking} and nonequilibrium network systems~\cite{94_liu2025dynamical,102_zhou2024k}. Kinetically constrained quantum dynamics~\cite{2lxs-wccj} is another promising direction that may be explored by quantum generative models~\cite{s6zj-vzdp}.




\section*{Acknowledgments}
We acknowledge Online Club Nanothermodynamica for helpful discussions. This work is supported by Project 12322501, 12575035, 12325501, 12247104, 12447101, 12405047 of National Natural Science Foundation of China, and 2026NSFSCZY0124 of the Natural Science Foundation of Sichuan Province. 
P.Z. is supported by the Strategic Priority Research Program of the Chinese Academy of Sciences, Grant No. XDB1680000.  
The HPC is supported by the Center for HPC at University of Electronic Science and Technology of China.

\section*{Data availability}
The authors declare that the data supporting this study are available within the paper.
A PyTorch code implementation of the present algorithm is openly available on GitHub [to be inserted upon acceptance]. 

\appendix

\section*{Appendix}

\begin{figure}
\includegraphics[width=0.7\columnwidth]{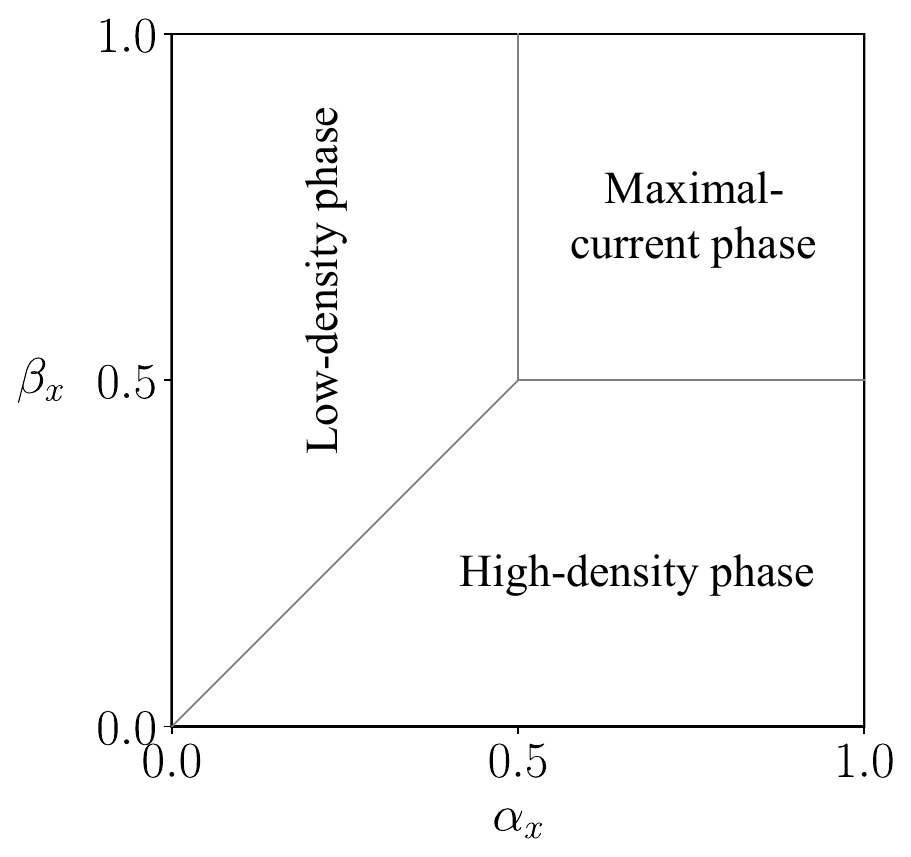}
\caption{\label{fig:1DTASEP_phase_graph}\textbf{Phase diagram of the 1D TASEP.} The maximal-current (MC) phase occurs for $\alpha_x > 1/2$ and $\beta_x > 1/2$.
The low-density (LD) phase is found when $\alpha_x < 1/2$ and $\beta_x > \alpha_x$.
The high-density (HD) phase corresponds to $\beta_x < 1/2$ and $\alpha_x > \beta_x$.
The line $\alpha_x = \beta_x < 1/2$ marks a first-order phase transition between the LD and HD phases.
}
\end{figure}
\begin{table*}[htbp]
  \centering
  \caption{Comparison of the VAN estimates of the mean density ($\rho$) and mean current ($J$) at $L=20$, $s=0$, and $t=10^3$ with the exact values of the open 1D TASEP~\cite{31_Derrida1992,31_derrida1993exact,schutz1993phase}. Here, $\alpha_x$ and $\beta_x$ denote the particle injection and extraction rates, respectively.}
  \label{tab:table2}
  \begin{tabular*}{\textwidth}{@{\extracolsep{\fill}} c c c c c c c @{}}
      \toprule
      \makecell[b]{Physical \\ quantity} &
      \makecell[b]{Phase \\ type} &
      \makecell[b]{Boundary \\ conditions} &
      \makecell[b]{Theoretical \\ result} &
      \makecell[b]{Exemplified \\ rates ($\alpha_x, \beta_x$)} &
      \makecell[b]{Theoretical \\ value} &
      \makecell[b]{VAN \\ result} \\
      \midrule
      \multirow{3}{*}{Steady-state density}
      & Low-density     & $\alpha_x < \beta_x, \alpha_x < 1/2$ & $\rho = \alpha_x$     & $(0.2, 0.8)$ & 0.20 & 0.201 \\
      & High-density    & $\alpha_x > \beta_x, \beta_x < 1/2$  & $\rho = 1 - \beta_x$  & $(0.8, 0.2)$ & 0.80 & 0.800 \\
      & Maximal-current & $\alpha_x, \beta_x \ge 1/2$          & $\rho = 0.5$           & $(0.8, 0.8)$ & 0.50 & 0.498 \\
      \midrule
      \multirow{3}{*}{Current}
      & Low-density     & $\alpha_x < \beta_x, \alpha_x < 1/2$ & \multirow{3}{*}{$J = \rho(1-\rho)$} & $(0.2, 0.8)$ & 0.16 & 0.160 \\
      & High-density    & $\alpha_x > \beta_x, \beta_x < 1/2$  &                                      & $(0.8, 0.2)$ & 0.16 & 0.160 \\
      & Maximal-current & $\alpha_x, \beta_x \ge 1/2$          &                                      & $(0.8, 0.8)$ & 0.25 & 0.253 \\
      \bottomrule
  \end{tabular*}
\end{table*}
\begin{figure*}
\includegraphics[width=1\textwidth]{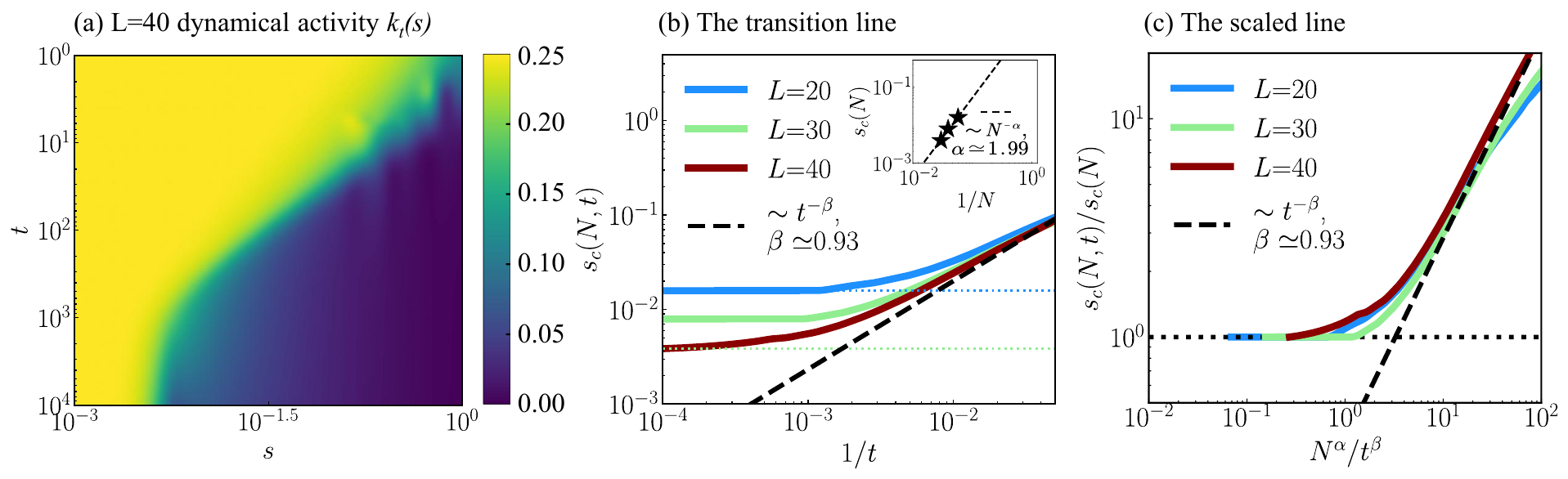}
\caption{\label{fig:1DSSEP}\textbf{The dynamical active-inactive phase transition of 1D SSEP.} (a) Heatmap of the dynamical activity $k_t(s)$ for a system of size $L = 40$, illustrating its evolution across the counting field $s$ and trajectory time $t$, where the colorbar indicates the activity magnitude. (b) Finite-time critical counting field $s_c(N, t)$ as a function of inverse time $1/t$ for system sizes $L \in \{20, 30, 40\}$. The dashed line represents a power-law fit $s_c(N, t)-s_c(N) \sim t^{-\beta}$ with the dynamical exponent $\beta \approx 0.93$. Inset: Steady-state critical field $s_c(N) \sim N^{-\alpha}$ with $\alpha \approx 1.97$. (c) Finite-time scaling collapse of $s_c(N, t)/s_c(N)$ plotted against the rescaled time $N^\alpha/t^\beta$ using the exponents $\alpha \approx 1.97$ and $\beta \approx 0.93$, indicating the proper scaling relation.
}
\end{figure*}
\section{Principal notation and conventions}\label{app:notation}

For reference, the principal notation and conventions used throughout the paper are summarized in Table~\ref{tab:notation}.

\section{1D TASEP density and current benchmark}\label{app:1d-tasep-steady-state}

We benchmark the expressive capacity of the variational autoregressive network (VAN) using the open 1D totally asymmetric simple exclusion process (TASEP), whose unbiased steady-state density and current are known exactly from the matrix product ansatz (MPA)~\cite{31_Derrida1992,31_derrida1993exact}. The VAN calculations use $L=20$, a counting field $s=0$, and trajectory time $t=10^3$. In this unbiased, long-time regime, the mean observables can be compared directly with the exact steady-state results. Fig.~\ref{fig:1DTASEP_phase_graph} shows the reference phase diagram, and Table~\ref{tab:table2} compares the VAN estimates with the exact values.

We test representative low-density, high-density, and maximal-current boundary conditions. Across these regimes, the VAN estimates are close to the corresponding exact phase-dependent values of the density and current, with maximum relative deviations of approximately $0.5\%$ for the mean density and $1.2\%$ for the mean current. Because the calculation uses finite $L$ and $t$, exact equality with the thermodynamic-limit steady-state values is not expected. Their close agreement shows that the VAN can represent the unbiased boundary-driven distribution and recover its steady-state observables, supporting its use in the subsequent finite-time and higher-dimensional calculations.

\section{1D SSEP benchmark}\label{app:1D SSEP benchmark}

We benchmark the finite-time expressive capacity of the VAN against matrix-product-state (MPS) results for the 1D symmetric simple exclusion process~\cite{12_causer2022finite}. The activity map for $L=40$ captures the active-inactive crossover as a function of $s$ and $t$ (Fig.~\ref{fig:1DSSEP}(a)). The critical field follows $s_c(N,t)-s_c(N)\sim t^{-\beta}$ with $\beta\approx0.93$, while its long-time value scales as $s_c(N)\sim N^{-\alpha}$ with $\alpha\approx1.97$ (Fig.~\ref{fig:1DSSEP}(b)). Using these exponents, the data for $L=20$ and $L=40$ collapse onto a common curve (Fig.~\ref{fig:1DSSEP}(c)), consistent with the MPS finite-time scaling results.

Agreement at the levels of the activity landscape, transition location, and finite-time and finite-size scaling shows that the VAN can represent the evolving tilted distribution throughout the finite-time regime, rather than only its long-time limit. This benchmark supports extending the same variational representation to the higher-dimensional systems considered below.
\section{Natural-gradient versus Adam optimization}\label{app:optimizer-comparison}
\begin{figure*}
\includegraphics[width=1\textwidth]{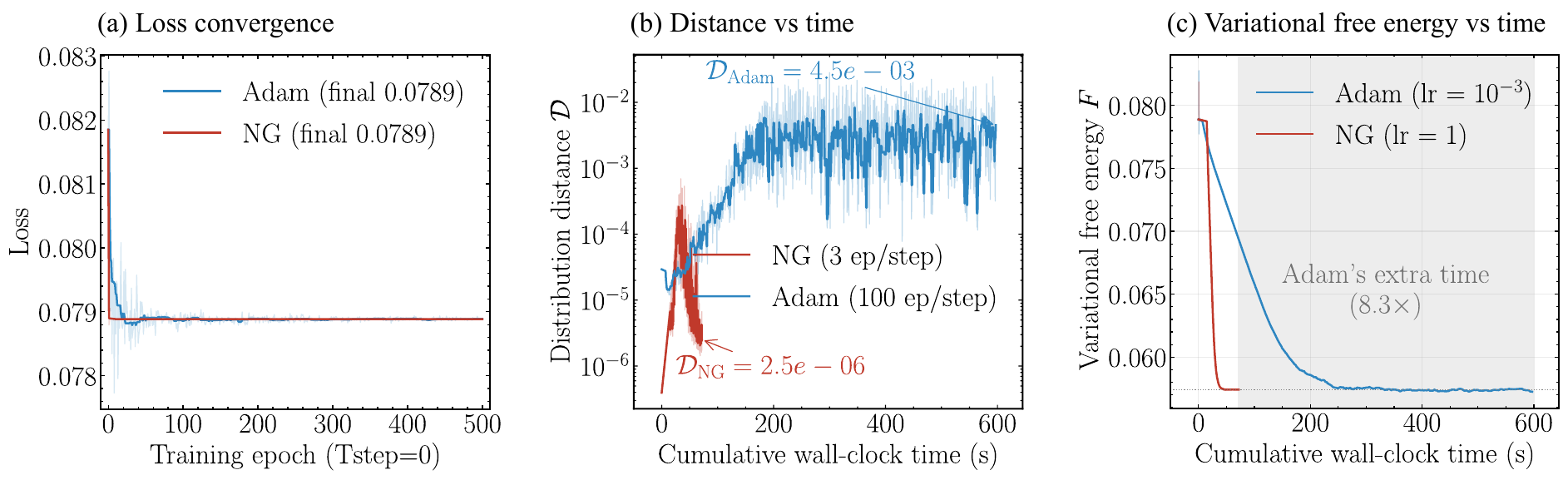}
\caption{\label{fig:Loss}\textbf{Comparison of natural-gradient (NG) and Adam optimization for VAN training of the 2D SSEP} (L=3, seed=42, identical initialization). (a) Training loss on a fixed target \(P\). Both methods converge to a final loss of 0.0789, with NG reaching convergence in 2 epochs compared with 68 epochs for Adam. (b) Distribution distance (Euclidean distance) \(D\) over the 1001-step evolution as a function of cumulative wall-clock time, with NG and Adam using 3 and 100 epochs per step, respectively. NG reaches \(D=2.5\times10^{-6}\), whereas Adam saturates at \(D=4.5\times10^{-3}\), corresponding to an approximately 1800-fold gap. (c) Training loss versus cumulative wall-clock time. NG reaches the final loss in 1.2 min, compared with 10.0 min for Adam, corresponding to an 8.3× speed-up; the shaded region indicates the additional time required by Adam.
}
\end{figure*}
Fig.~\ref{fig:Loss} motivates our use of natural-gradient optimization in the finite-time evolution. For a fixed target distribution, NG and Adam reach the same final loss, but NG converges in substantially fewer epochs, indicating that its probability-space geometry provides a more effective update direction. This advantage becomes more important when the optimization is repeated at every time step: small fitting errors can accumulate over the evolution, and Adam retains a much larger distribution error even with many more epochs per step. Thus, although an individual NG epoch is more expensive because it requires constructing and solving the Fisher-matrix system, the reduction in the required number of epochs yields both higher accuracy and a lower total wall-clock cost. We therefore use blocked NG throughout the finite-time calculations.

\section{Mean-field estimate of the 1D activity}\label{app:mean-field-activity}

We derive the mean-field expression used to interpret the activity ordering in Fig.~\ref{fig:1DASEP_boundary_effect_merge}. Let $\rho_i=\langle n_i\rangle$ be the local density on a chain of length $L$. The mean instantaneous number of bulk hops per unit time, plus the boundary insertion and removal events, is approximated by
\begin{align}
k_{\mathrm{MF}}&=k_{\mathrm{bulk}}^{\mathrm{MF}}
+k_{\mathrm{bdry}}^{\mathrm{MF}},\label{eq:mf-activity-general}\\
k_{\mathrm{bulk}}^{\mathrm{MF}}
&=\frac{1}{L}\sum_{i=1}^{L-1}\Big[
p_x\rho_i(1-\rho_{i+1})
+q_x\rho_{i+1}(1-\rho_i)\Big],\\
k_{\mathrm{bdry}}^{\mathrm{MF}}
&=\frac{1}{L}\Big[\alpha_x(1-\rho_1)+\gamma_x\rho_1+\delta_x(1-\rho_L)+\beta_x\rho_L\Big],
\end{align}
where the factorization $\langle n_i(1-n_j)\rangle\simeq\rho_i(1-\rho_j)$ neglects nearest-neighbor correlations. For the 1D calculations in the main text, $p_x+q_x=1$ and the boundary contribution is of order $L^{-1}$. Neglecting this subextensive term and assuming that the density varies slowly between neighboring sites gives
\begin{equation}
k_{\mathrm{MF}}\simeq \frac{1}{L}\sum_{i=1}^{L}\rho_i(1-\rho_i).
\label{eq:mf-activity-local}
\end{equation}
Defining $\bar\rho=L^{-1}\sum_i\rho_i$ and $\mathrm{Var}(\rho)=L^{-1}\sum_i(\rho_i-\bar\rho)^2$, we obtain
\begin{equation}
\begin{split}
k_{\mathrm{MF}}
&\simeq \bar\rho-\left[\bar\rho^2+\mathrm{Var}(\rho)\right]
=\frac{1}{4}-\left(\bar\rho-\frac{1}{2}\right)^2-\mathrm{Var}(\rho).
\end{split}
\label{eq:mf-activity-density}
\end{equation}
Thus, within this approximation, activity is favored by a mean density near one-half and by a spatially homogeneous density profile. Because Eq.~\eqref{eq:mf-activity-density} neglects correlations, boundary terms, density gradients, and the modification of the effective dynamics by a finite counting field, we use it only as a qualitative estimate in the weakly tilted regime $s=10^{-2}$, rather than as an exact expression for $k(s)$.

\section{Directional-density criterion for 2D TASEP}\label{app:2d-directional-convergence}
This appendix gives the mean-field argument behind the directional-density mapping used in Sec.~\ref{subsec:Transverse-longitudinal flow coupling in two-dimensional system}, following earlier mean-field treatments of coupled and 2D exclusion processes~\cite{37_ding2019mean}. Consider a 2D TASEP on an $L\times L$ square lattice with open boundaries along both directions. The $x$ direction is controlled by $(\alpha_x,\beta_x)$ and the $y$ direction by $(\alpha_y,\beta_y)$. For the homogeneous phases, let $\rho_{\mathrm{1D}}(\alpha,\beta)$ denote the bulk density selected by the exact 1D open-boundary TASEP phase diagram~\cite{31_Derrida1992,31_derrida1993exact}:
\begin{equation}
\rho_{\mathrm{1D}}(\alpha,\beta)=
\begin{cases}
\alpha, & \alpha<\beta,\ \alpha<1/2,\\
1-\beta, & \beta<\alpha,\ \beta<1/2,\\
1/2, & \alpha\geq1/2,\ \beta\geq1/2.
\end{cases}
\end{equation}
The directional-density matching condition is
\begin{equation}
\rho_{\mathrm{1D}}(\alpha_x,\beta_x)
=\rho_{\mathrm{1D}}(\alpha_y,\beta_y)
\equiv \rho^*.
\end{equation}
This construction extends directly to a $d$-dimensional hypercubic lattice by requiring $\rho_{\mathrm{1D}}(\alpha_\mu,\beta_\mu)=\rho^*$ for every direction $\mu=1,\ldots,d$. We validate it here only for $d=2$, where $\rho^*$ fixes the leading bulk density up to finite boundary layers. The coexistence line, where the 1D state contains a shock rather than a homogeneous bulk density, is treated separately.

\begin{table}[t]
\centering 
\caption{\label{tab:2dtasep-numerical-MFA}Finite-size mean-field estimates of the density ($\hat\rho$) and directional currents ($\hat J_x,\hat J_y$) of the 2D TASEP. In each numerical entry, the first and second lines give the results for $L=5$ and $L=30$, respectively, and the third line in parentheses gives the bulk prediction inherited from the corresponding 1D boundary-selected phase. For both finite sizes, the outermost one-site boundary layer is excluded when calculating the spatially averaged density and currents. For coexistence, $\bar\rho$ denotes the spatially averaged density.}
\footnotesize
\begin{tabular}{ccc}
\toprule
\makecell[c]{Phase \\ type} & \makecell[c]{Density \\ $\hat\rho$ ($\rho$)} & \makecell[c]{Currents \\ $\hat J_x,\hat J_y$ ($J_x,J_y$)} \\
\midrule
\makecell[c]{Low-density\\$\alpha_{x(y)}=0.1$, \\ $\alpha_{x(y)}<\beta_{x(y)}$} &
\makecell[c]{$\hat\rho=0.103$ \\ $\hat\rho=0.101$ \\ $(\rho=0.100)$} &
\makecell[c]{$\hat J_x=0.090,\hat J_y=0.092$ \\ $\hat J_x=0.090,\hat J_y=0.091$ \\ $(J_x=0.090,J_y=0.090)$}  \\
\addlinespace[0.5em]

\makecell[c]{Coexistence \\ $\alpha_{x} = \beta_{x}=0.4$ \\ $\alpha_{y} = \beta_{y}=0.4$} &
\makecell[c]{$\bar\rho=0.500$ \\ $\bar\rho=0.500$ \\ $(\bar\rho=0.500)$} &
\makecell[c]{$\hat J_x=0.233,\hat J_y=0.233$ \\ $\hat J_x=0.240,\hat J_y=0.240$ \\ $(J_x=0.240,J_y=0.240)$}  \\
\addlinespace[0.5em]

\makecell[c]{Maximal-current \\ $\alpha_{x},\beta_{x} \ge 0.5$ \\ $\alpha_{y},\beta_{y} \ge 0.5$} &
\makecell[c]{$\hat\rho=0.481$ \\ $\hat\rho=0.493$ \\ $(\rho=0.500)$} &
\makecell[c]{$\hat J_x=0.262,\hat J_y=0.277$ \\ $\hat J_x=0.249,\hat J_y=0.251$ \\ $(J_x=0.250,J_y=0.250)$} \\
\addlinespace[0.5em]
\makecell[c]{High-density\\$\beta_{x(y)}=0.3$ \\$\alpha_{x(y)}>\beta_{x(y)}$} &
\makecell[c]{$\hat\rho=0.662$ \\ $\hat\rho=0.690$ \\ $(\rho=0.700)$} &
\makecell[c]{$\hat J_x=0.208,\hat J_y=0.217$ \\ $\hat J_x=0.210,\hat J_y=0.213$ \\ $(J_x=0.210,J_y=0.210)$} \\
\bottomrule
\end{tabular}
\end{table}

At $L=5$, the mean-field results agree closely with the VAN estimates in Table~\ref{tab:2dtasep-numerical}, with maximum absolute discrepancies of $0.008$ in the density and $0.007$ in the directional currents, despite the VAN values being evaluated at the weak tilt $s=10^{-3}$ and finite time $t=50$. Increasing the mean-field system size to $L=30$ further reduces the finite-size deviations: the low-density, maximal-current, and high-density densities become $0.101$, $0.493$, and $0.690$, approaching the predictions $0.100$, $0.500$, and $0.700$, respectively, while the mean density in the coexistence regime remains $0.500$. The maximal-current pair likewise approaches $(J_x,J_y)=(0.250,0.250)$, changing from $(0.262,0.277)$ at $L=5$ to $(0.249,0.251)$ at $L=30$. Because the values at both sizes exclude the outermost one-site boundary layer, these results support finite-size convergence of the interior density and currents toward the directional-density predictions.
\begin{figure*}
\includegraphics[width=1\textwidth]{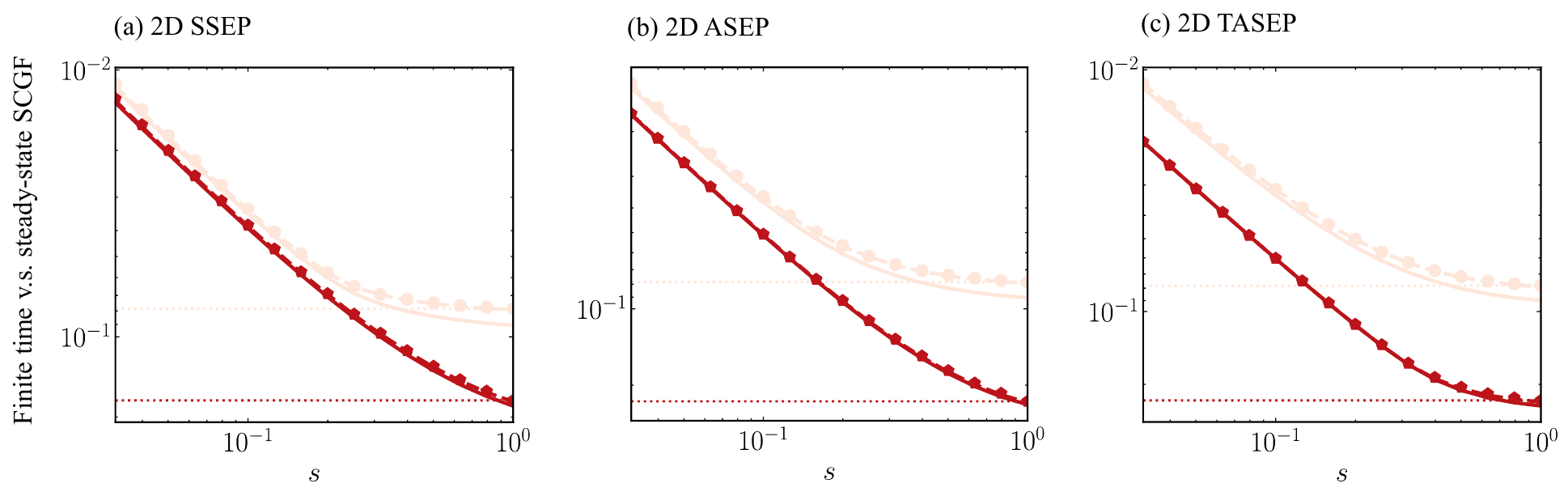}
\caption{\label{fig:Finite time evolution v.s. SCGF}\textbf{Finite-time evolution toward the steady state in 2D SEPs.} The SCGF $\psi(s)/L^2$ as a function of the counting field $s$ is compared across (a) 2D SSEP, (b) 2D ASEP, and (c) 2D TASEP. Solid lines show the finite-time results at $t=50$, while dashed lines show long-time estimates obtained by extrapolating the evolution to $t=135$. Markers (circles for low activity and pentagons for high activity) denote steady-state VMC results. Their agreement with the dashed curves indicates that the long-time extrapolation provides an accurate estimate of the steady-state SCGF for the tested cases. Results are shown for the boundary rates $(\alpha_x,\beta_x,\alpha_y,\beta_y)=(0.2,0.8,0.2,0.8)$ (light, low activity) and $(0.6,0.6,0.6,0.6)$ (dark red, high activity).
}
\end{figure*}
\subsection{Homogeneous low-density phase}
Suppose both directions are in the low-density phase with equal injection rates $\alpha_x=\alpha_y$, where $\alpha_x<1/2$ and $\alpha_x<\beta_x,\beta_y$. The common target state is the 1D low-density bulk, $\rho_{\mathrm{bulk}}=\alpha_x$, while the exit rates may remain asymmetric. Away from the boundaries, the steady-state mean-field continuity equation is
\begin{equation}
\begin{split}
\rho_{x-1,y}(1-\rho_{x,y}) + \rho_{x,y-1}(1-\rho_{x,y}) \\
= \rho_{x,y}(1-\rho_{x+1,y}) + \rho_{x,y}(1-\rho_{x,y+1}).
\end{split}
\label{eq:2d-mf-continuity}
\end{equation}
Substituting the homogeneous profile $\rho_{x,y}=\alpha_x$ gives both sides equal to $2\alpha_x(1-\alpha_x)$, so the 1D-selected low-density state is an exact interior mean-field solution.

The effect of asymmetric extraction rates is confined to boundary layers. For example, at the right boundary $x=L-1$, away from corners, take the adjacent interior density as $\rho_{L-2,y}=\alpha_x$ and use translational invariance along $y$ to write $\rho_{L-1,y}=\rho_b$. The boundary continuity equation is
\begin{equation}
\alpha_x(1-\rho_b)+\rho_b(1-\rho_b)
=\beta_x\rho_b+\rho_b(1-\rho_b).
\end{equation}
The transverse self-coupling terms cancel, leaving
\begin{equation}
\rho_b=\frac{\alpha_x}{\alpha_x+\beta_x}.
\label{eq:ld-boundary-density}
\end{equation}
Thus the asymmetric exit rate changes the local boundary density but not the bulk density.

To estimate the boundary-layer width, write $\rho_{x,y}=\alpha_x+\epsilon_{x,y}$ and, away from corners, take $\epsilon_{x,y}=\epsilon_x$. Linearizing Eq.~\eqref{eq:2d-mf-continuity} gives
\begin{equation}
(1-\alpha_x)\epsilon_{x-1}-\epsilon_x+\alpha_x\epsilon_{x+1}=0.
\label{eq:ld-linearized}
\end{equation}
The characteristic roots are $\lambda_1=(1-\alpha_x)/\alpha_x>1$ and $\lambda_2=1$. The constant mode associated with $\lambda_2=1$ is excluded by matching to the bulk condition $\epsilon_x\to0$ away from the exit. Near the exit, using the inward distance $n=L-1-x$, the decaying solution is
\begin{equation}
\epsilon_n=(\rho_b-\alpha_x)\left(\frac{\alpha_x}{1-\alpha_x}\right)^n.
\end{equation}
The decay length is $\xi=-1/\ln[\alpha_x/(1-\alpha_x)]$, independent of $L$, so the boundary layer occupies a vanishing fraction of the system as $L\to\infty$. Corner effects break the 1D boundary invariance locally, but they occupy only an $\mathcal{O}(1)$ region and do not modify the bulk convergence.

\subsection{Homogeneous high-density phase}

The high-density case is the particle-hole counterpart. Suppose both directions are in the high-density phase with equal extraction rates $\beta_x=\beta_y$, where $\beta_x<1/2$ and $\beta_x<\alpha_x,\alpha_y$. The target bulk density is $\rho_{\mathrm{bulk}}=1-\beta_x$, while the injection rates may be asymmetric. The homogeneous profile $\rho_{x,y}=1-\beta_x$ satisfies Eq.~\eqref{eq:2d-mf-continuity}. At the left entrance boundary, the same cancellation of transverse self-coupling terms gives the local entrance density
\begin{equation}
\rho_{\mathrm{in}}=\frac{\alpha_x}{\alpha_x+\beta_x}.
\end{equation}
The perturbation away from this boundary decays exponentially into the bulk over an $\mathcal{O}(1)$ length scale. Hence asymmetric injection rates modify only boundary layers and do not change the macroscopic high-density bulk selected by the common exit rate.

\begin{figure*}
\includegraphics[width=1\textwidth]{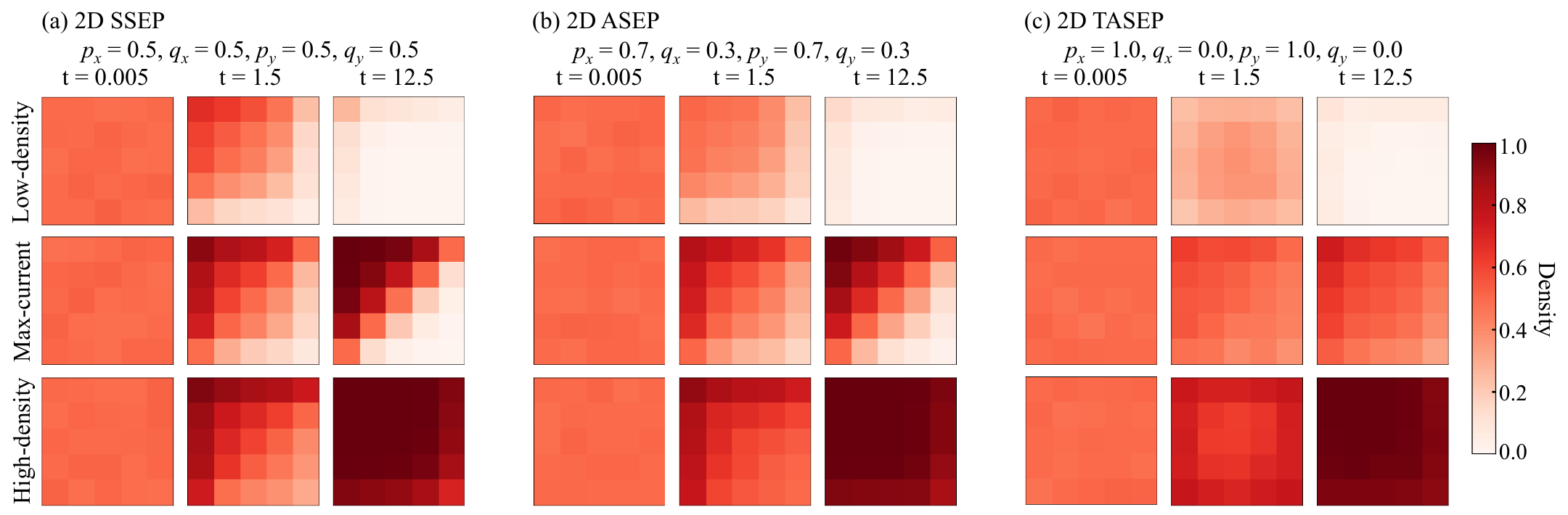}
\caption{\label{fig:2_2DASEP_spatial_structure}\textbf{Evolution of density profiles toward the long-time limit in 2D SEPs under different density regimes.} The columns show transient spatial density profiles at $s = 10^0$ for a system of size $L = 5$ for (a) SSEP, (b) ASEP, and (c) TASEP, with the corresponding bulk hopping rates $(p_x, q_x, p_y, q_y)$ indicated above each panel. The rows correspond to three density regimes set by the boundary coupling rates $(\alpha_x, \beta_x; \alpha_y, \beta_y)$: the low-density regime with $(0.2, 0.8; 0.2, 0.8)$ (top), the maximal-current regime with $(0.6, 0.6; 0.6, 0.6)$ (middle), and the high-density regime with $(0.8, 0.2; 0.8, 0.2)$ (bottom). For each driving condition, spatial density snapshots are shown at $t = 0.005$, $t = 1.5$, and $t = 12.5$. At $t = 12.5$, the density profiles show little visible change on subsequent time scales, indicating that the density structures used in the main-text $t=50$ profiles have effectively stabilized, even though the SCGF is extrapolated to $t=135$. The color bar denotes the local particle density.
}
\end{figure*}
\subsection{Maximal-current phase}

When $\alpha_x,\alpha_y\geq1/2$ and $\beta_x,\beta_y\geq1/2$, both directions select the maximal-current state with $\rho_{\mathrm{bulk}}=1/2$. Substitution into Eq.~\eqref{eq:2d-mf-continuity} satisfies the interior equation. Linearizing around $\rho_{x,y}=1/2+\epsilon_{x,y}$ and again taking $\epsilon_{x,y}=\epsilon_x$ away from corners yields
\begin{equation}
\epsilon_{x-1}-2\epsilon_x+\epsilon_{x+1}=0.
\end{equation}
The characteristic root is degenerate, $\lambda=1$, so the simple mean-field theory does not produce an exponentially localized boundary layer. This marginality is analogous to the algebraic boundary relaxation of the 1D maximal-current phase~\cite{31_derrida1993exact}, but does not by itself determine the boundary scaling in two dimensions. The VAN density profiles remain consistent with a bulk density near $1/2$ while resolving finite-size boundary structure beyond this simple mean-field description.

\begin{figure*}
\includegraphics[width=0.99\textwidth]{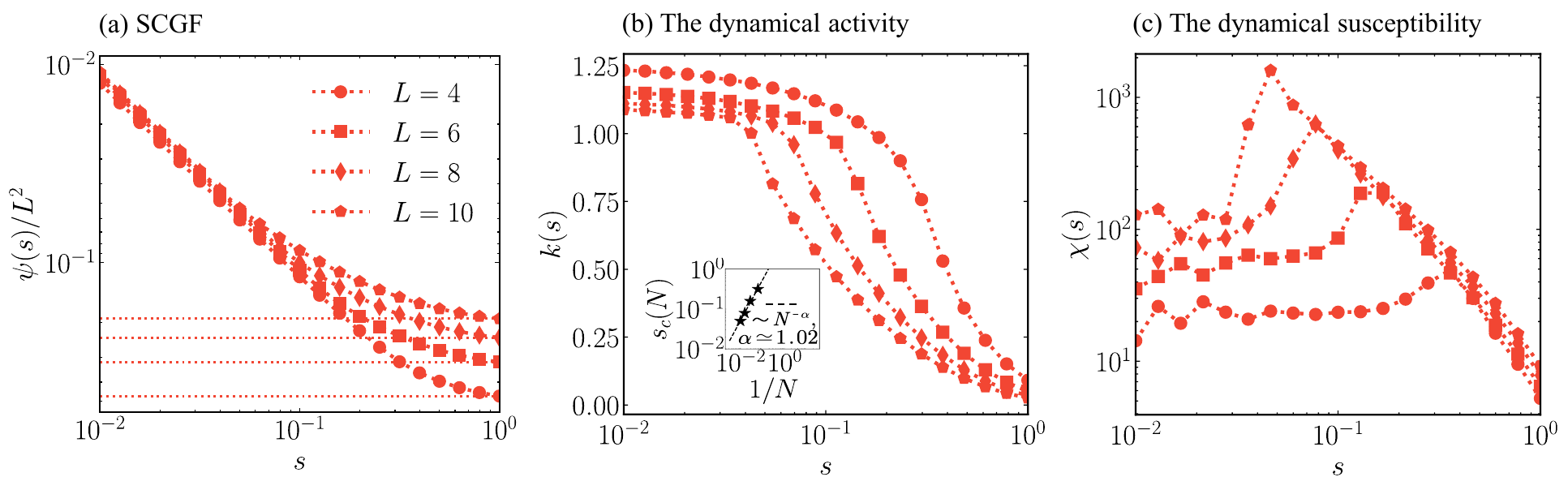}
\caption{\label{fig:comparison result2_1s}\textbf{Dynamical large deviations and active-inactive transitions in 2D SSEP via VMC.} (a) Steady-state scaled cumulant generating function (SCGF) $\psi(s)/L^2$ calculated via variational Monte Carlo (VMC) as a function of the counting field $s$ for system sizes $L \in \{4, 6, 8, 10\}$ (where $N = L^2$). The horizontal dotted lines show the value for $s\rightarrow \infty $. (b) Steady-state dynamical activity $k(s)$ versus $s$ for various $L$ in (a), showing a sharp crossover from an active phase to an inactive phase. Inset: Finite-size scaling of the critical field $s_c(N)$ as a function of the inverse system size $1/N$, fitted by the power law $s_c(N) \sim N^{-\alpha}$ with $\alpha \approx 1.02$. (c) Dynamical susceptibility $\chi(s) =\psi''(s)$ plotted against $s$. The pronounced peak in $\chi(s)$ grows and sharpens with increasing system size $L$, consistent with an emerging dynamical transition.
}
\end{figure*}

\begin{figure*}
\includegraphics[width=1\textwidth]{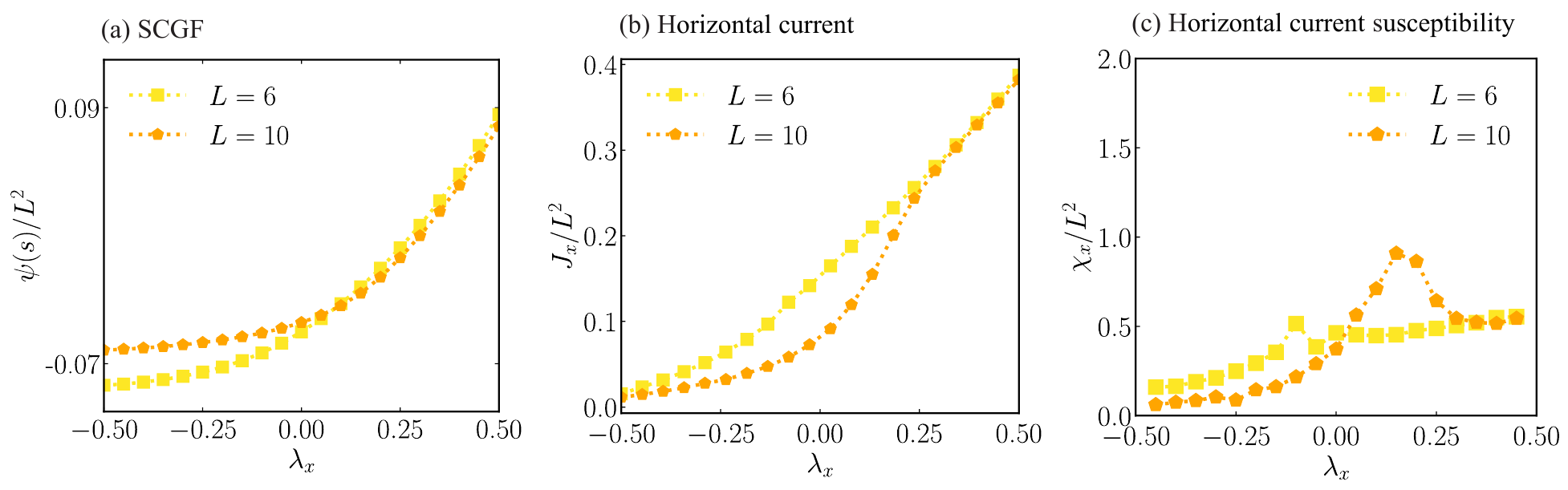}
\caption{\label{fig:comparison-result2-2s}\textbf{Current dynamical phase transitions in the 2D ASEP obtained with a VAN.} VAN results for a line cut through the dynamical phase space of the 2D ASEP with bulk hopping rates $p_{x(y)} = 1 - q_{x(y)} = 0.9$ and open boundary rates $\alpha_{x(y)} = \beta_{x(y)} = \gamma_{x(y)} = \delta_{x(y)} = 0.5$. From left to right, we show the per-site SCGF $\psi(\lambda_x, \lambda_y)/L^2$, horizontal current $J_x/L^2$, and horizontal current susceptibility $\chi_{x}/L^2$ at $\lambda_y = -0.5$ with $\lambda_x \in [-0.5, 0.5]$. Each line corresponds to a system size $L \in \left\{6, 10\right\}$.
}
\end{figure*}

\begin{figure*}
\includegraphics[width=1\textwidth]{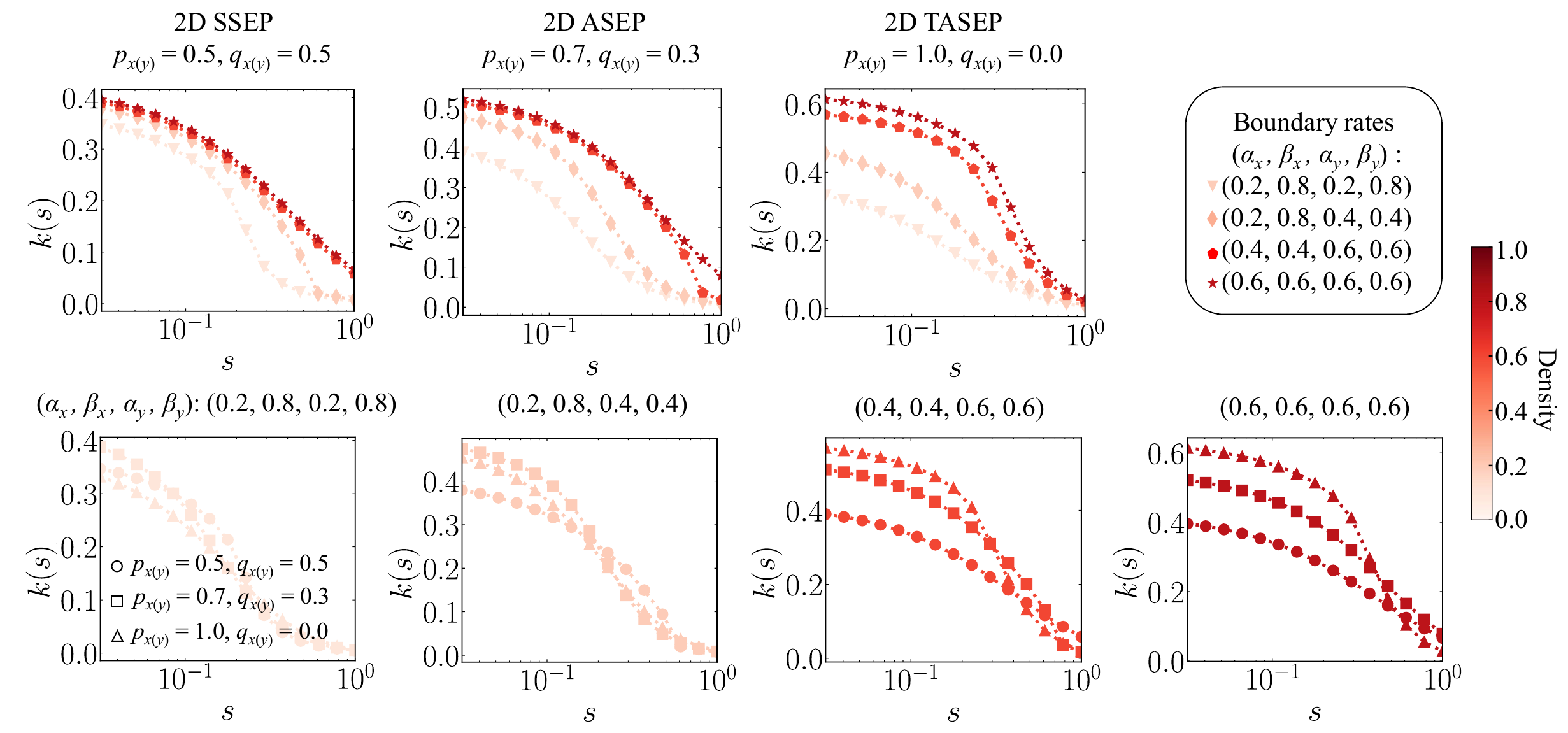}
\caption{\label{fig:2DSEP_activity_long-time}\textbf{Boundary and bulk effects on the steady-state dynamical activity of 2D SEPs.} (Top row) Steady-state dynamical activity $k(s)$ as a function of the counting field $s$ for $L=5$ in the 2D SSEP [$p_{x(y)}=q_{x(y)}=0.5$, left], ASEP [$p_{x(y)}=0.7$, $q_{x(y)}=0.3$, middle], and TASEP [$p_{x(y)}=1.0$, $q_{x(y)}=0.0$, right]. Colors denote four sets of boundary rates $(\alpha_x,\beta_x,\alpha_y,\beta_y)$, ranging from low-activity (light) to high-activity (dark red) regimes. (Bottom row) Effect of bulk driving on $k(s)$ for the same four sets of boundary rates, with the hopping asymmetry corresponding to the SSEP (circles), ASEP (squares), and TASEP (triangles).
}
\end{figure*}
\subsection{Coexistence line}

On the coexistence line, $\alpha_x=\beta_x=\lambda<1/2$ and $\alpha_y=\beta_y=\lambda<1/2$, the 1D state is not characterized by a single homogeneous bulk density. Instead, a delocalized shock separates domains with densities $\lambda$ and $1-\lambda$. Extending the directional-density criterion to this shared coexistence structure therefore suggests a phase-separated 2D profile rather than homogeneous convergence. The two limiting density regions satisfy
\begin{equation}
\rho_{\mathrm{injection\ side}}\to\lambda,
\qquad
\rho_{\mathrm{exit\ side}}\to1-\lambda.
\end{equation}
In the symmetric square geometry, this construction is consistent with a diagonal shock-like interface between the injection-dominated and exit-dominated regions. The criterion fixes the two limiting densities but not the interface position, width, or finite-size fluctuations, which remain properties of the full 2D dynamics.
\begin{figure*}
\includegraphics[width=1\textwidth]{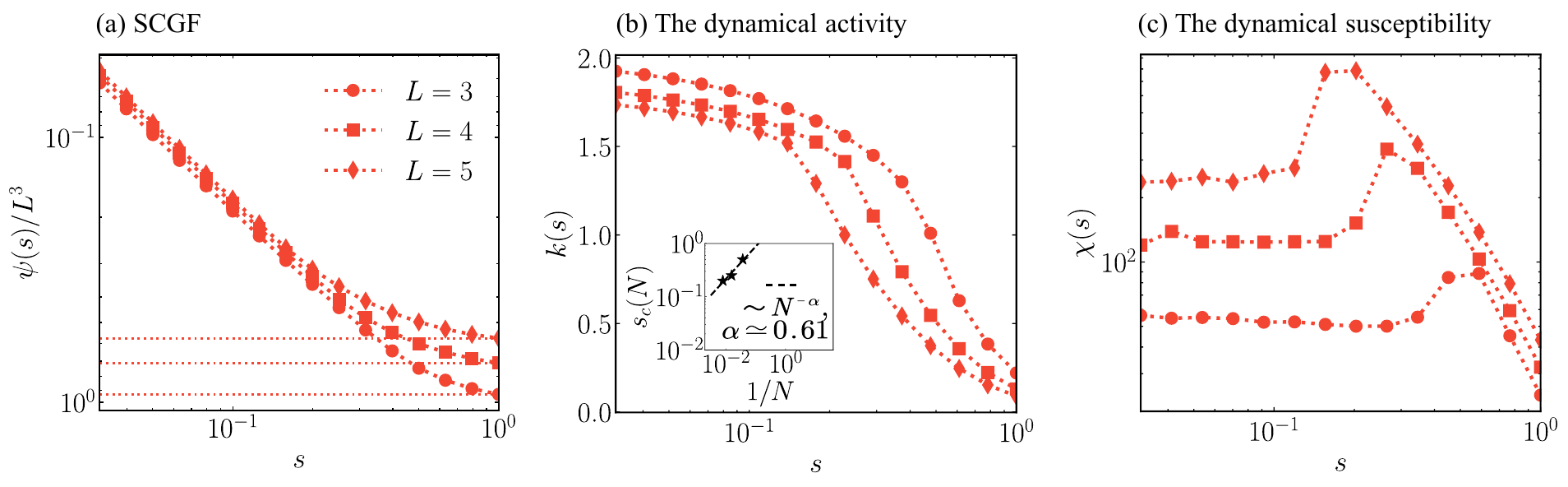}
\caption{\label{fig:comparison-3dresult-1s}\textbf{Dynamical large deviations and active-inactive transitions in 3D SSEP via VMC.} (a) Steady-state scaled cumulant generating function (SCGF) $\psi(s)/L^3$ calculated via variational Monte Carlo (VMC) as a function of the counting field $s$ for system sizes $L \in \{3, 4, 5\}$ (where $N = L^3$). The horizontal dotted lines show the value for $s \rightarrow \infty$. (b) Steady-state dynamical activity $k(s)$ versus $s$ for various $L$, showing a sharp crossover from an active phase to an inactive phase. Inset: Finite-size scaling of the critical field $s_c(N)$ as a function of the inverse system size $1/N$, fitted by the power law $s_c(N) \sim N^{-\alpha}$ with $\alpha \approx 0.61$. (c) Dynamical susceptibility $\chi(s) = \psi''(s)$ plotted against $s$. The pronounced peak in $\chi(s)$ grows and sharpens with increasing system size $L$, consistent with an emerging dynamical transition.
}
\end{figure*}
\begin{figure*}
\includegraphics[width=1\textwidth]{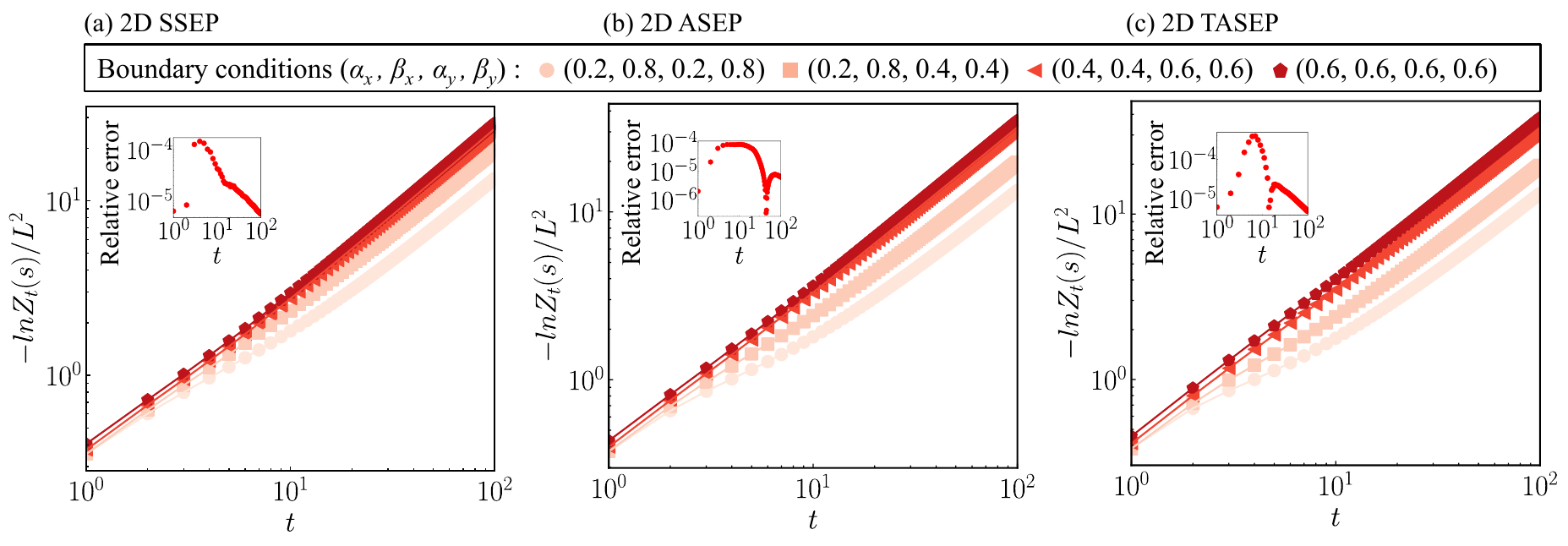}
\caption{\label{fig:tracking-time-evolution-dynamical-partition-function}\textbf{Tracking the time evolution of the dynamical partition function in 2D SEPs and its relative error.} The panels show the scaled negative logarithmic dynamical partition function $-\ln Z_t(s)/L^2$ as a function of trajectory time $t$ at $s = 10^0$ for a system of size $L = 3$ under (a) 2D SSEP, (b) 2D ASEP, and (c) 2D TASEP. The curves correspond to four representative boundary coupling conditions $(\alpha_x, \beta_x, \alpha_y, \beta_y)$ representing different activity regimes. Insets: Relative error of the variational network results compared with the numerically exact solutions obtained by tracking the full probability distribution. The maximum relative error remains on the order of $10^{-4}$ throughout the evolution, showing that the neural-network-based variational framework captures the tested transient nonequilibrium dynamics with controlled numerical error.
}
\end{figure*}

\begin{figure*}
\includegraphics[width=1\textwidth]{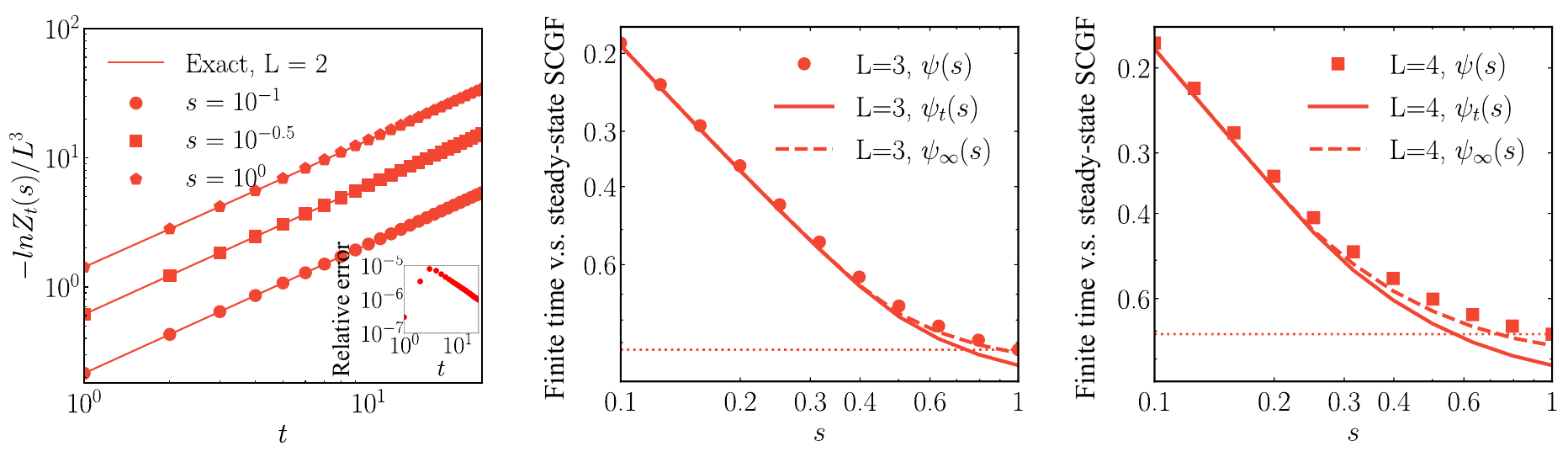}
\caption{\label{fig:3d-validation-exact-solutions}\textbf{Accuracy and finite-time evolution toward the steady state in the 3D SSEP.} Left panel: Scaled negative logarithm of the dynamical partition function, $-\ln Z_t(s)/L^3$, as a function of trajectory time $t$ for $L=2$ at counting fields $s=10^{-1}$, $10^{-0.5}$, and $10^0$. The inset shows the relative error of the VAN results with respect to exact numerical solutions at $s=10^{0}$, which remains below $10^{-5}$. Middle and right panels: SCGF as a function of the counting field $s$ for $L=3$ (middle) and $L=4$ (right). Solid lines show the finite-time results $\psi_t(s)$ at $t=25$, whereas dashed lines show the long-time estimates $\psi_\infty(s)$ obtained by extrapolating to $t=1200$. Markers (circles for $L=3$ and squares for $L=4$) indicate the steady-state SCGF $\psi(s)$ obtained using variational Monte Carlo (VMC), and dotted horizontal lines indicate the $s\to\infty$ values. The agreement between the long-time estimates and steady-state VMC results confirms convergence to the nonequilibrium steady state.
}
\end{figure*}
\begin{table*}[htbp]
  \centering
  \caption{Summary of VAN hyperparameters and computational cost.}
  \label{tab:van_fullwidth}
  \newcommand{\midbox}[1]{\makebox[0.14\textwidth][c]{#1}}
  \begin{tabular*}{\textwidth}{@{\extracolsep{\fill}}ccccc@{}}
    \toprule
    \multicolumn{1}{c}{\raisebox{2.5ex}{\makecell[c]{ Calculation type}}} &
    \multicolumn{3}{c}{\makebox[0.42\textwidth][c]{\makecell[c]{Finite-time evolution: \\ first time step uses about 500 epochs, \\ later time steps use 3 epochs}}} &
    \multicolumn{1}{c}{\raisebox{1.5ex}{\makecell[c]{Steady-state VMC: \\ 1200 epochs}}} \\
    \midrule
    \makecell[c]{Dimension} & \midbox{\makecell[c]{1}} & \midbox{\makecell[c]{2}} & \midbox{\makecell[c]{3}} & \makecell[c]{2} \tabularnewline[5pt]
    
    \makecell[c]{Time steps} & \midbox{\makecell[c]{$2\times10^4$}} & \midbox{\makecell[c]{$10^4$}} & \midbox{\makecell[c]{$5\times10^3/5\times10^3/2\times10^3$}} & \makecell[c]{--} \tabularnewline[5pt]
    \makecell[c]{VAN type} & \midbox{\makecell[c]{MADE}} & \midbox{\makecell[c]{MADE}} & \midbox{\makecell[c]{MADE/MADE/Transformer}} & \makecell[c]{MADE} \tabularnewline[5pt]
    \begin{tabular}[c]{@{}c@{}}VAN architecture\\ (depth, width)\end{tabular} & \midbox{\makecell[c]{$(4, 16)$}} & \midbox{\makecell[c]{$(4, 16)$}} & \midbox{\makecell[c]{$(4, 16)/(4,16)$/Transformer: Appendix~\ref{app:Computational cost}}} & \makecell[c]{$(4, 16)$} \tabularnewline[10pt]
    \makecell[c]{Lattice size $L$} & \midbox{\makecell[c]{20}} & \midbox{\makecell[c]{$4/6/8$}} & \midbox{\makecell[c]{$3/4/5$}} & \makecell[c]{$4/6/8/10$} \tabularnewline[5pt]
    \begin{tabular}[c]{@{}c@{}}Wall-clock time\\ per $s$ value (hour)\end{tabular} & \midbox{\makecell[c]{$1.47$}} & \midbox{\makecell[c]{$1.97/6.8/24.5$}} & \midbox{\makecell[c]{$0.90/1.96/9.64$}} & \makecell[c]{$0.05/0.29/0.47/1.65$} \tabularnewline
    \bottomrule
  \end{tabular*}
  \vspace{2mm}
\end{table*}
\section{2D SSEP and ASEP benchmarks}\label{app:2d-ssep-benchmark}
We first assess the finite-time expressive capacity of the VAN in two dimensions. Fig.~\ref{fig:Finite time evolution v.s. SCGF} compares the finite-time SCGF at $t=50$, its long-time extrapolation to $t=135$, and independent steady-state VMC results for the SSEP, ASEP, and TASEP. The agreement between the extrapolated curves and the VMC markers shows that the VAN evolution captures the approach to the steady-state SCGF for the tested boundary conditions. Fig.~\ref{fig:2_2DASEP_spatial_structure} further tracks the density-profile evolution across different driving and density regimes. The profiles change substantially at early times but show little visible variation after $t=12.5$, indicating that the spatial density structure stabilizes well before the SCGF reaches its extrapolated long-time limit.

For the steady-state 2D SSEP, the VAN reproduces the SCGF, dynamical activity, susceptibility peak, and critical-field scaling $s_c(N)\sim N^{-\alpha}$ with $N=L^2$ (Fig.~\ref{fig:comparison result2_1s}). The fitted exponent $\alpha\approx1.02$ is close to the tensor-network estimate $\alpha\approx0.92$ reported in~\cite{19_causer2023optimal} for the finite system sizes considered. These results show that the autoregressive ansatz can represent the tilted-ensemble observables and the active-inactive crossover of a 2D diffusive system.

For the driven ASEP, the VAN estimates of the SCGF, horizontal current, and current susceptibility along the selected line in counting-field space are broadly consistent with PEPS calculations~\cite{18_helms2020dynamical} (Fig.~\ref{fig:comparison-result2-2s}). Together, these finite-time and steady-state benchmarks demonstrate that the VAN can represent two-dimensional tilted ensembles and resolve their dynamical phase-transition signatures under both diffusive and driven dynamics.

As a further complement to the susceptibility analysis, Fig.~\ref{fig:2DSEP_activity_long-time} shows the steady-state dynamical activity $k(s)$ under the same boundary and bulk variations. The top row compares different boundary-rate sets at fixed bulk dynamics, while the bottom row follows each fixed boundary-rate set as the bulk dynamics changes from the SSEP through the ASEP to the TASEP. These results provide the activity-level counterpart to the boundary- and bulk-dependent susceptibility trends in Fig.~\ref{fig:2DASEP_current_density_steady}.

\section{Steady state of 3D SSEP via VMC}\label{app:3d-ssep-vmc}

To complement the finite-time 3D results, we use variational Monte Carlo (VMC) to calculate the steady-state SCGF, dynamical activity and susceptibility for $L\in\{3,4,5\}$. As shown in Fig.~\ref{fig:comparison-3dresult-1s}, the susceptibility peak grows and sharpens with increasing system size, and the critical field follows $s_c(N)\sim N^{-\alpha}$ with $\alpha\approx0.61$. This steady-state estimate is close to the finite-time result $\alpha\approx0.60$ obtained in the main text, providing a comparison while indicating finite-time and finite-size corrections over the accessible system sizes.

\section{Tracking the dynamical partition function}\label{app:finite-time-partition-validation}

We validate the finite-time VAN evolution by tracking the dynamical partition function against exact numerical results. For the 2D SSEP, ASEP, and TASEP with $L=3$, the relative error remains on the order of $10^{-4}$ throughout the tested evolution and across different boundary conditions (Fig.~\ref{fig:tracking-time-evolution-dynamical-partition-function}). The 3D SSEP benchmark gives an error below $10^{-5}$ for $L=2$ (Fig.~\ref{fig:3d-validation-exact-solutions}). For the larger 3D systems with $L=3$ and $L=4$, the same figure shows that the long-time SCGF extrapolated from the finite-time evolution is broadly consistent with the optimized steady-state VMC results. Together, these comparisons demonstrate that the VAN accurately represents the finite-time tilted distribution in the exactly accessible systems and can be consistently extrapolated toward the steady state.
\section{Computational cost}\label{app:Computational cost}

Table~\ref{tab:van_fullwidth} summarizes the VAN architecture and the wall-clock time used for both finite-time evolution and steady-state variational Monte Carlo (VMC) calculations. The reported computational time is measured for a single value of the counting field $s$. We consider one-, two-, and three-dimensional systems with different lattice lengths, and use different numbers of time steps for the finite-time calculations. The time step is $\delta t=0.05$ in 1D and $\delta t=0.005$ in both 2D and 3D. All calculations use a batch size of $N_B=1000$, a natural-gradient learning rate $\eta=1$, and a damping parameter $\lambda_d=10^{-5}$. For finite-time evolution, the first time step is optimized for about 500 epochs, while subsequent time steps are updated for $3$ epochs. The 3D $L=5$ calculation instead uses a Transformer with token dimension $4$, positional dimension $124$, two layers, and four attention heads, and is evolved for $2000$ time steps. For the steady-state VMC calculation, we list the 2D computational cost used for comparison with the long-time limit of the present finite-time method. All timings were obtained on a single H100 GPU. Multiple entries in the lattice-size, time-step, architecture, and wall-clock rows correspond to the listed values of $L$ in the same column.

\bibliography{apssamp}

@article{1_derrida2007non,
title = {Non-equilibrium steady states: fluctuations and large deviations of the density and of the current},
author = {Derrida, Bernard},
journal = {J. Stat. Mech.},
volume = {2007},
number = {07},
pages = {P07023},
year = {2007},
publisher = {IOP Publishing},
doi = {10.1088/1742-5468/2007/07/P07023}
}

@article{2_Chou_2011,
title = {Non-equilibrium statistical mechanics: from a paradigmatic model to biological transport},
author = {Chou, T. and Mallick, K. and Zia, R. K. P.},
journal = {Rep. Prog. Phys.},
volume = {74},
number = {11},
pages = {116601},
year = {2011},
doi = {10.1088/0034-4885/74/11/116601},
url = {https://dx.doi.org/10.1088/0034-4885/74/11/116601}
}

@article{4_krug1997origins,
title = {Origins of scale invariance in growth processes},
author = {Krug, Joachim},
journal = {Adv. Phys.},
volume = {46},
number = {2},
pages = {139--282},
year = {1997},
publisher = {Taylor & Francis},
doi = {10.1080/00018739700101498}
}

@article{6_schadschneider2000statistical,
title = {Statistical physics of traffic flow},
author = {Schadschneider, Andreas},
journal = {Physica A},
volume = {285},
number = {1--2},
pages = {101--120},
year = {2000},
publisher = {Elsevier},
doi = {10.1016/S0378-4371(00)00274-0}
}

@article{8_touchette2009large,
title = {The large deviation approach to statistical mechanics},
author = {Touchette, Hugo},
journal = {Phys. Rep.},
volume = {478},
number = {1--3},
pages = {1--69},
year = {2009},
publisher = {Elsevier},
doi = {10.1016/j.physrep.2009.05.002}
}

@article{9_garrahan2009first,
title = {First-order dynamical phase transition in models of glasses: an approach based on ensembles of histories},
author = {Garrahan, Juan P. and Jack, Robert L. and Lecomte, Vivien and Pitard, Estelle and van Duijvendijk, Kristina and van Wijland, Fr{\'e}d{\'e}ric},
journal = {J. Phys. A: Math. Theor.},
volume = {42},
number = {7},
pages = {075007},
year = {2009},
publisher = {IOP Publishing},
doi = {10.1088/1751-8113/42/7/075007}
}

@article{10_gorissen2012exact,
title = {Exact current statistics of the asymmetric simple exclusion process with open boundaries},
author = {Gorissen, Mieke and Lazarescu, Alexandre and Mallick, Kirone and Vanderzande, Carlo},
journal = {Phys. Rev. Lett.},
volume = {109},
number = {17},
pages = {170601},
year = {2012},
publisher = {APS},
doi = {10.1103/PhysRevLett.109.170601}
}

@article{11_derrida1998exact,
title = {Exact large deviation function in the asymmetric exclusion process},
author = {Derrida, Bernard and Lebowitz, Joel L.},
journal = {Phys. Rev. Lett.},
volume = {80},
number = {2},
pages = {209},
year = {1998},
publisher = {APS},
doi = {10.1103/PhysRevLett.80.209}
}

@article{12_causer2022finite,
title = {Finite time large deviations via matrix product states},
author = {Causer, Luke and Ba{\~n}uls, Mari Carmen and Garrahan, Juan P.},
journal = {Phys. Rev. Lett.},
volume = {128},
number = {9},
pages = {090605},
year = {2022},
publisher = {APS},
doi = {10.1103/PhysRevLett.128.090605}
}

@article{13_ray2018importance,
title = {Importance sampling large deviations in nonequilibrium steady states. I},
author = {Ray, Ushnish and Chan, Garnet Kin and Limmer, David T},
journal = {J. Chem. Phys.},
volume = {148},
number = {12},
pages = {124120},
year = {2018},
publisher = {AIP Publishing},
doi = {10.1063/1.5003151}
}

@article{14_golinelli2006asymmetric,
title = {The asymmetric simple exclusion process: an integrable model for non-equilibrium statistical mechanics},
author = {Golinelli, Olivier and Mallick, Kirone},
journal = {J. Phys. A: Math. Gen.},
volume = {39},
number = {41},
pages = {12679},
year = {2006},
publisher = {IOP Publishing},
doi = {10.1088/0305-4470/39/41/S03}
}

@article{15_carleo2017solving,
title = {Solving the quantum many-body problem with artificial neural networks},
author = {Carleo, Giuseppe and Troyer, Matthias},
journal = {Science},
volume = {355},
number = {6325},
pages = {602--606},
year = {2017},
doi = {10.1126/science.aag2302}
}

@article{16_wu2019solving,
title = {Solving statistical mechanics using variational autoregressive networks},
author = {Wu, Dian and Wang, Lei and Zhang, Pan},
journal = {Phys. Rev. Lett.},
volume = {122},
number = {8},
pages = {080602},
year = {2019},
publisher = {APS},
doi = {10.1103/PhysRevLett.122.080602}
}

@article{17_tang2024learning,
title = {Learning nonequilibrium statistical mechanics and dynamical phase transitions},
author = {Tang, Ying and Liu, Jing and Zhang, Jiang and Zhang, Pan},
journal = {Nat. Commun.},
volume = {15},
number = {1},
pages = {1117},
year = {2024},
doi = {10.1038/s41467-024-45172-8}
}

@article{18_helms2020dynamical,
  title={Dynamical phase transitions in a 2D classical nonequilibrium model via 2D tensor networks},
  author={Helms, Phillip and Chan, Garnet Kin-Lic},
  journal={Phys. Rev. Lett.},
  volume={125},
  number={14},
  pages={140601},
  year={2020},
  publisher={APS},
  doi={10.1103/PhysRevLett.125.140601}
}

@article{19_causer2023optimal,
title = {Optimal sampling of dynamical large deviations in two dimensions via tensor networks},
author = {Causer, Luke and Ba{\~n}uls, Mari Carmen and Garrahan, Juan P.},
journal = {Phys. Rev. Lett.},
volume = {130},
number = {14},
pages = {147401},
year = {2023},
publisher = {APS},
doi = {10.1103/PhysRevLett.130.147401}
}

@article{20_Liu2025Efficient,
title = {Efficient optimization of variational autoregressive networks with natural gradient},
author = {Liu, Jing and Tang, Ying and Zhang, Pan},
journal = {Phys. Rev. E},
volume = {111},
issue = {2},
pages = {025304},
numpages = {11},
year = {2025},
month = {Feb},
publisher = {American Physical Society},
doi = {10.1103/PhysRevE.111.025304},
url = {https://link.aps.org/doi/10.1103/PhysRevE.111.025304}
}

@inproceedings{21_Germain2015MADE,
title = {MADE: Masked Autoencoder for Distribution Estimation},
author = {Germain, Mathieu and Gregor, Karol and Murray, Iain and Larochelle, Hugo},
booktitle = {Proc. 32nd Int. Conf. Mach. Learn.},
volume = {37},
pages = {881--889},
year = {2015},
publisher = {PMLR},
url = {http://proceedings.mlr.press/v37/germain15.html}
}

@article{22_luo2022autoregressive,
title = {Autoregressive neural network for simulating open quantum systems via a probabilistic formulation},
author = {Luo, Di and Chen, Zhuo and Carrasquilla, Juan and Clark, Bryan K.},
journal = {Phys. Rev. Lett.},
volume = {128},
number = {9},
pages = {090501},
year = {2022},
doi = {10.1103/PhysRevLett.128.090501}
}

@article{23_casert2021dynamical,
title = {Dynamical large deviations of two-dimensional kinetically constrained models using a neural-network state ansatz},
author = {Casert, Corneel and Vieijra, Tom and Whitelam, Stephen and Tamblyn, Isaac},
journal = {Phys. Rev. Lett.},
volume = {127},
number = {12},
pages = {120602},
year = {2021},
doi = {10.1103/PhysRevLett.127.120602}
}

@article{24_amari1998natural,
title = {Natural gradient works efficiently in learning},
author = {Amari, Shun-ichi},
journal = {Neural Comput.},
volume = {10},
number = {2},
pages = {251--276},
year = {1998},
month = {02},
doi = {10.1162/089976698300017746}
}

@incollection{27_schmittmann1995statistical,
title = {Statistical mechanics of driven diffusive systems},
author = {B. Schmittmann and R.K.P. Zia},
booktitle = {Phase Transit. Crit. Phenom.},
volume = {17},
pages = {3--214},
year = {1995},
publisher = {Academic Press}
}

@article{28_spitzer1970interaction,
title = {Interaction of Markov processes},
author = {Spitzer, Frank},
journal = {Adv. Math.},
volume = {5},
number = {2},
pages = {246--290},
year = {1970},
doi = {10.1016/0001-8708(70)90034-4}
}

@article{29_widom1991repton,
title = {Repton model of gel electrophoresis and diffusion},
author = {Widom, B. and Viovy, J. L. and Defontaines, A. D.},
journal = {J. Phys. I},
volume = {1},
number = {12},
pages = {1759--1784},
year = {1991},
doi = {10.1051/jp1:1991239}
}

@article{31_Derrida1992,
  author  = {Derrida, B. and Domany, E. and Mukamel, D.},
  title   = {An exact solution of a one-dimensional asymmetric exclusion model with open boundaries},
  journal = {J. Stat. Phys.},
  volume  = {69},
  number  = {3},
  pages   = {667--687},
  year    = {1992},
  doi     = {10.1007/BF01050430}
}

@article{31_derrida1993exact,
title = {Exact solution of a 1D asymmetric exclusion model using a matrix formulation},
author = {Derrida, Bernard and Evans, Martin R. and Hakim, Vincent and Pasquier, Vincent},
journal = {J. Phys. A: Math. Gen.},
volume = {26},
number = {7},
pages = {1493},
year = {1993},
doi = {10.1088/0305-4470/26/7/011}
}

@article{32_derrida2003exact,
title = {Exact large deviation functional of a stationary open driven diffusive system: the asymmetric exclusion process},
author = {Derrida, B. and Lebowitz, J. L. and Speer, E. R.},
journal = {J. Stat. Phys.},
volume = {110},
number = {3},
pages = {775--810},
year = {2003},
doi = {10.1023/A:1022111919402}
}

@article{37_ding2019mean,
title = {Mean-field analysis for Asymmetric Exclusion Processes on two parallel lattices with fully parallel dynamics},
journal = {Physica A},
volume = {516},
pages = {317--326},
year = {2019},
issn = {0378-4371},
doi = {10.1016/j.physa.2018.09.167},
url = {https://www.sciencedirect.com/science/article/pii/S0378437118312895},
author = {Ding, Zhongjun and Liu, Teng and Lou, Xinxin and Shen, Zhiwei and Zhu, Kongjin and Jiang, Rui and Wang, Binghong and Chen, Bokui},

}

@article{41_ray2018exact,
title = {Exact fluctuations of nonequilibrium steady states from approximate auxiliary dynamics},
author = {Ray, Ushnish and Chan, Garnet Kin-Lic and Limmer, David T.},
journal = {Phys. Rev. Lett.},
volume = {120},
number = {21},
pages = {210602},
year = {2018},
doi = {10.1103/PhysRevLett.120.210602}
}

@article{42_perez2019sampling,
title = {Sampling rare events across dynamical phase transitions},
author = {P{\'e}rez-Espigares, Carlos and Hurtado, Pablo I.},
journal = {Chaos},
volume = {29},
number = {8},
pages = {083106},
year = {2019},
doi = {10.1063/1.5091669}
}

@article{43_donsker1975asymptotic,
title = {Asymptotic evaluation of certain Markov process expectations for large time, I},
author = {Donsker, Monroe D. and Varadhan, S. R. S. Srinivasa},
journal = {Commun. Pure Appl. Math.},
volume = {28},
number = {1},
pages = {1--47},
year = {1975},
doi = {10.1002/cpa.3160280102}
}

@article{45_nemoto2017finite,
title = {Finite-size scaling of a first-order dynamical phase transition: Adaptive population dynamics and an effective model},
author = {Nemoto, Takahiro and Jack, Robert L. and Lecomte, Vivien},
journal = {Phys. Rev. Lett.},
volume = {118},
number = {11},
pages = {115702},
year = {2017},
doi = {10.1103/PhysRevLett.118.115702}
}

@article{47_klymko2018rare,
title = {Rare behavior of growth processes via umbrella sampling of trajectories},
author = {Klymko, Katherine and Geissler, Phillip L. and Garrahan, Juan P. and Whitelam, Stephen},
journal = {Phys. Rev. E},
volume = {97},
number = {3},
pages = {032123},
year = {2018},
doi = {10.1103/PhysRevE.97.032123}
}

@article{48_jacobson2019direct,
title = {Direct evaluation of dynamical large-deviation rate functions using a variational ansatz},
author = {Jacobson, Daniel and Whitelam, Stephen},
journal = {Phys. Rev. E},
volume = {100},
number = {5},
pages = {052139},
year = {2019},
doi = {10.1103/PhysRevE.100.052139}
}

@article{50_de2011large,
title = {Large deviation function for the current in the open asymmetric simple exclusion process},
author = {de Gier, Jan and Essler, Fabian H. L.},
journal = {Phys. Rev. Lett.},
volume = {107},
number = {1},
pages = {010602},
year = {2011},
doi = {10.1103/PhysRevLett.107.010602}
}

@article{51_gorissen2011finite,
title = {Finite size scaling of current fluctuations in the totally asymmetric exclusion process},
author = {Gorissen, Mieke and Vanderzande, Carlo},
journal = {J. Phys. A: Math. Theor.},
volume = {44},
number = {11},
pages = {115005},
year = {2011},
doi = {10.1088/1751-8113/44/11/115005}
}

@article{54_brewer2018efficient,
title = {Efficient characterisation of large deviations using population dynamics},
author = {Brewer, Tobias and Clark, Stephen R. and Bradford, Russell and Jack, Robert L.},
journal = {J. Stat. Mech.},
volume = {2018},
number = {5},
pages = {053204},
year = {2018},
doi = {10.1088/1742-5468/aab3ef}
}

@article{63_hibat2020recurrent,
title = {Recurrent neural network wave functions},
author = {Hibat-Allah, Mohamed and Ganahl, Martin and Hayward, Lauren E. and Melko, Roger G. and Carrasquilla, Juan},
journal = {Phys. Rev. Res.},
volume = {2},
number = {2},
pages = {023358},
year = {2020},
doi = {10.1103/PhysRevResearch.2.023358}
}

@article{65_garrahan2007dynamical,
title = {Dynamical first-order phase transition in kinetically constrained models of glasses},
author = {Garrahan, Juan P. and Jack, Robert L. and Lecomte, Vivien and Pitard, Estelle and van Duijvendijk, Kristina and van Wijland, Fr{\'e}d{\'e}ric},
journal = {Phys. Rev. Lett.},
volume = {98},
number = {19},
pages = {195702},
year = {2007},
doi = {10.1103/PhysRevLett.98.195702}
}

@article{68_banuls2019using,
title = {Using matrix product states to study the dynamical large deviations of kinetically constrained models},
author = {Ba{\~n}uls, Mari Carmen and Garrahan, Juan P.},
journal = {Phys. Rev. Lett.},
volume = {123},
number = {20},
pages = {200601},
year = {2019},
doi = {10.1103/PhysRevLett.123.200601}
}

@article{69_causer2021optimal,
title = {Optimal sampling of dynamical large deviations via matrix product states},
author = {Causer, Luke and Ba{\~n}uls, Mari Carmen and Garrahan, Juan P.},
journal = {Phys. Rev. E},
volume = {103},
number = {6},
pages = {062144},
year = {2021},
doi = {10.1103/PhysRevE.103.062144}
}

@article{73_rose2021reinforcement,
title = {A reinforcement learning approach to rare trajectory sampling},
author = {Rose, Dominic C. and Mair, Jamie F. and Garrahan, Juan P.},
journal = {New J. Phys.},
volume = {23},
number = {1},
pages = {013013},
year = {2021},
doi = {10.1088/1367-2630/abd7bd}
}

@article{78_ha2003queuing,
title = {Queuing transitions in the asymmetric simple exclusion process},
author = {Ha, Meesoon and Timonen, Jussi and den Nijs, Marcel},
journal = {Phys. Rev. E},
volume = {68},
number = {5},
pages = {056122},
year = {2003},
publisher = {American Physical Society},
doi = {10.1103/PhysRevE.68.056122},
url = {https://doi.org/10.1103/PhysRevE.68.056122}
}

@article{79_derrida1997shock,
title = {Shock profiles for the asymmetric simple exclusion process in one dimension},
author = {Derrida, Bernard and Lebowitz, Joel L. and Speer, Eugene R.},
journal = {J. Stat. Phys.},
volume = {89},
number = {1--2},
pages = {135--167},
year = {1997},
publisher = {Springer},
doi = {10.1007/BF02770758},
url = {https://doi.org/10.1007/BF02770758}
}

@article{weng2025tracking,
  title={Tracking large chemical reaction networks and rare events by neural networks},
  author={Weng, Jiayu and Zhu, Xinyi and Liu, Jing and L{\"u}, Linyuan and Zhang, Pan and Tang, Ying},
  journal={arXiv:2512.10309},
  year={2025},
  url={https://arxiv.org/abs/2512.10309}
}

@article{80_maes2020frenesy,
title = {Frenesy: Time-symmetric dynamical activity in nonequilibria},
author = {Maes, Christian},
journal = {Phys. Rep.},
volume = {850},
pages = {1--33},
year = {2020},
doi = {10.1016/j.physrep.2020.01.002}
}

@article{81_lecomte2012inactive,
doi = {10.1088/1751-8113/45/17/175001},
url = {https://doi.org/10.1088/1751-8113/45/17/175001},
year = {2012},
month = {apr},
publisher = {IOP Publishing},
volume = {45},
number = {17},
pages = {175001},
author = {Lecomte, Vivien and Garrahan, Juan P. and van Wijland, Fr{\'e}d{\'e}ric},
title = {Inactive dynamical phase of a symmetric exclusion process on a ring},
journal = {J. Phys. A: Math. Theor.}
}

@article{82_vanicat2021mapping,
author = {Vanicat, Matthieu and Bertin, Eric and Lecomte, Vivien and Ragoucy, Eric},
year = {2021},
month = {02},
pages = {028},
title = {Mapping current and activity fluctuations in exclusion processes: consequences and open questions},
volume = {10},
journal = {SciPost Phys.},
doi = {10.21468/SciPostPhys.10.2.028}
}

@article{84_Bertini2015Macroscopic,
  title = {Macroscopic fluctuation theory},
  author = {Bertini, Lorenzo and De Sole, Alberto and Gabrielli, Davide and Jona-Lasinio, Giovanni and Landim, Claudio},
  journal = {Rev. Mod. Phys.},
  volume = {87},
  issue = {2},
  pages = {593--636},
  numpages = {44},
  year = {2015},
  month = {Jun},
  publisher = {American Physical Society},
  doi = {10.1103/RevModPhys.87.593},
  url = {https://link.aps.org/doi/10.1103/RevModPhys.87.593}
}

@article{85_lecomte2007thermodynamic,
  title={Thermodynamic formalism for systems with Markov dynamics},
  author={Lecomte, Vivien and Appert-Rolland, C{\'e}cile and Van Wijland, Fr{\'e}d{\'e}ric},
  journal={J. Stat. Phys.},
  volume={127},
  number={1},
  pages={51--106},
  year={2007},
  publisher={Springer},
  doi = {10.1007/s10955-006-9254-0}
}

@article{86_appert2008universal,
  title={Universal cumulants of the current in diffusive systems on a ring},
  author={Appert-Rolland, C{\'e}cile and Derrida, Bernard and Lecomte, Vivien and Van Wijland, Fr{\'e}d{\'e}ric},
  journal={Phys. Rev. E},
  volume={78},
  number={2},
  pages={021122},
  year={2008},
  publisher={APS},
  doi = {10.1103/PhysRevE.78.021122}
}

@article{87_bertini2005current,
  title={Current fluctuations in stochastic lattice gases},
  author={Bertini, Lorenzo and De Sole, Alberto and Gabrielli, Davide and Jona-Lasinio, Gianni and Landim, Claudio},
  journal={Phys. Rev. Lett.},
  volume={94},
  number={3},
  pages={030601},
  year={2005},
  publisher={APS},
  doi={10.1103/PhysRevLett.94.030601}
}

@article{88_kullback1951information,
title = {On Information and Sufficiency},
author = {Kullback, Solomon and Leibler, Richard A.},
journal = {Ann. Math. Stat.},
volume = {22},
number = {1},
pages = {79--86},
year = {1951},
doi = {10.1214/aoms/1177729694}
}

@article{89_williams1992reinforce,
title = {Simple Statistical Gradient-Following Algorithms for Connectionist Reinforcement Learning},
author = {Williams, Ronald J.},
journal = {Mach. Learn.},
volume = {8},
pages = {229--256},
year = {1992},
doi = {10.1007/BF00992696}
}

@inproceedings{90_kingma2015adam,
  title = {Adam: A Method for Stochastic Optimization},
  author = {Kingma, Diederik P. and Ba, Jimmy},
  booktitle = {Proc. Int. Conf. Learn. Represent.},
  year = {2015},
  eprint = {1412.6980},
  archivePrefix = {arXiv},
  primaryClass = {cs.LG}
}

@article{91_reh2021time,
  title = {Time-Dependent Variational Principle for Open Quantum Systems with Artificial Neural Networks},
  author = {Reh, Moritz and Schmitt, Markus and G\"arttner, Martin},
  journal = {Phys. Rev. Lett.},
  volume = {127},
  number = {23},
  pages = {230501},
  year = {2021},
  publisher = {American Physical Society},
  doi = {10.1103/PhysRevLett.127.230501}
}

@article{92_Chuanbo_2024Distilling,
author = {Liu, Chuanbo and Wang, Jin},
title = {Distilling dynamical knowledge from stochastic reaction networks},
journal = {Proc. Natl. Acad. Sci. U.S.A.},
volume = {121},
number = {14},
pages = {e2317422121},
year = {2024},
doi = {10.1073/pnas.2317422121},
}

@article{93_fang2023divide,
author = {Fang, Zhou and Gupta, Ankit and Kumar, Sant and Khammash, Mustafa},
title = {A divide-and-conquer method for analyzing high-dimensional noisy gene expression networks},
journal = {bioRxiv},
year = {2023},
doi = {10.1101/2022.10.28.514278}
}

@article{94_liu2025dynamical,
title = {Dynamical Phase Transitions in Nonequilibrium Networks},
author = {Liu, Jiazhen and Aden, Nathaniel M. and Sarker, Debasish and Song, Chaoming},
journal = {Phys. Rev. Lett.},
volume = {135},
number = {16},
pages = {167402},
year = {2025},
doi = {10.1103/yls4-kdvj}
}

@article{95_Melko2024Language,
    author = "Melko, Roger G. and Carrasquilla, Juan",
    title = "{Language models for quantum simulation}",
    doi = "10.1038/s43588-023-00578-0",
    journal = "Nat. Comput. Sci.",
    volume = "4",
    number = "1",
    pages = "11--18",
    year = "2024"
}

@article{25_chen2024empowering,
title = {Empowering deep neural quantum states through efficient optimization},
author = {Chen, Ao and Heyl, Markus},
journal = {Nat. Phys.},
volume = {20},
number = {9},
pages = {1476--1481},
year = {2024},
doi = {10.1038/s41567-024-02566-1}
}

@article{77_Rende2024large-scale,
author = {Rende, Riccardo and Viteritti, Luciano and Bardone, Lorenzo and Becca, Federico and Goldt, Sebastian},
title = {A simple linear algebra identity to optimize large-scale neural network quantum states},
journal = {Commun. Phys.},
volume = {7},
pages = {260},
year = {2024},
doi = {10.1038/s42005-024-01732-4}
}

@inproceedings{98_li2026weightflow,
  title={WeightFlow: Learning stochastic dynamics via evolving weight of neural network},
  author={Li, Ruikun and Liu, Jiazhen and Wang, Huandong and Liao, Qingmin and Li, Yong},
  booktitle={Proc. AAAI Conf. Artif. Intell.},
  volume={40},
  pages={641--649},
  year={2026},
  doi={10.1609/aaai.v40i1.37029} 
}

@article{99_cai2026revival,
  title={Revival of variational method in noisy cell signaling with Fourier observer},
  author={Cai, Ruobing and Lan, Yueheng},
  journal={Commun. Theor. Phys.}, 
  volume={78},
  number={1},
  pages={015601},
  year={2026},
  publisher={IOP Publishing},
  doi={10.1088/1572-9494/adf38c}
}

@article{100_pagare2024stochastic,
  title={Stochastic distinguishability of Markovian trajectories},
  author={Pagare, Asawari and Zhang, Zhongmin and Zheng, Jiming and Lu, Zhiyue},
  journal={J. Chem. Phys.}, 
  volume={160},
  number={17},
  pages={171101},
  year={2024},
  publisher={AIP Publishing},
  doi={10.1063/5.0203335} 
}

@article{101_bulgarelli2025flow,
  title={Flow-based sampling for entanglement entropy and the machine learning of defects},
  author={Bulgarelli, Andrea and Cellini, Elia and Jansen, Karl and K{\"u}hn, Stefan and Nada, Alessandro and Nakajima, Shinichi and Nicoli, Kim A and Panero, Marco},
  journal={Phys. Rev. Lett.}, 
  volume={134},
  number={15},
  pages={151601},
  year={2025},
  publisher={APS},
  doi={10.1103/PhysRevLett.134.151601} 
}

@article{102_zhou2024k,
  title={K-core attack, equilibrium K-core, and kinetically constrained spin system},
  author={Zhou, Hai-Jun},
  journal={Chin. Phys. B}, 
  volume={33},
  number={6},
  pages={066402},
  year={2024},
  publisher={Chinese Physical Society and IOP Publishing Ltd},
  doi={10.1088/1674-1056/ad4329}
}

@article{103_muzzi2024principal,
  title={Principal component analysis of absorbing state phase transitions},
  author={Muzzi, Cristiano and Cortes, Ronald Santiago and Bhakuni, Devendra Singh and Jeli{\'c}, Asja and Gambassi, Andrea and Dalmonte, Marcello and Verdel, Roberto},
  journal={Phys. Rev. E}, 
  volume={110},
  number={6},
  pages={064121},
  year={2024},
  publisher={APS},
  doi={10.1103/PhysRevE.110.064121} 
}

@article{104_xiong2025capturing,
  title={Capturing the dynamics of the phase transition of skyrmions with a nonstationary machine learning approach},
  author={Xiong, Long and Zhou, Neng-Ji and Hu, Shi-Qian and Nian, Lei-Lei and Zheng, Bo},
  journal={Phys. Rev. B}, 
  volume={111},
  number={18},
  pages={184415},
  year={2025},
  publisher={APS},
  doi={10.1103/PhysRevB.111.184415} 
}

@misc{105_suzuki2025machine,
  title={Machine learning topological defect formation},
  author={Suzuki, Fumika and Li, Ying Wai and Zurek, Wojciech H.},
  year={2025},
  eprint={2508.20347},
  archivePrefix={arXiv},
  primaryClass={cond-mat.stat-mech}
}

@article{106_zhu2026two,
  title={Two-stage dynamics and phase control of skyrmion formation in chiral magnets},
  author={Zhu, Shiwei and Guan, Xinyuan and Sun, Zhen and Zhang, Qiuyao and Song, Changsheng},
  journal={Phys. Rev. B}, 
  volume={113},
  number={10},
  pages={L100403},
  year={2026},
  publisher={APS},
  doi={10.1103/z6k3-1zvy}
}

@article{108_miedema2017correlation,
  title = {Correlation Imaging Reveals Specific Crowding Dynamics of Kinesin Motor Proteins},
  author = {Miedema, Daniel M. and Kushwaha, Vandana S. and Denisov, Dmitry V. and Acar, Seyda and Nienhuis, Bernard and Peterman, Erwin J. G. and Schall, Peter},
  journal = {Phys. Rev. X},
  volume = {7},
  number = {4},
  pages = {041037},
  year = {2017},
  doi = {10.1103/PhysRevX.7.041037}
}

@article{109_stutzer2026stochastic,
  title = {Stochastic Calculus for Pathwise Observables of Markov-Jump Processes: Unification of Diffusion and Jump Dynamics},
  author = {Stutzer, Lars Torbj{\o}rn and Dieball, Cai and Godec, Alja{\v{z}}},
  journal = {Phys. Rev. X},
  volume = {16},
  number = {2},
  pages = {021038},
  year = {2026},
  doi = {10.1103/9ncx-4pgr}
}

@article{110_DERRIDA199865,
title = {An exactly soluble non-equilibrium system: The asymmetric simple exclusion process},
journal = {Phys. Rep.},
volume = {301},
number = {1},
pages = {65--83},
year = {1998},
issn = {0370-1573},
doi = {10.1016/S0370-1573(98)00006-4},
author = {B. Derrida}
}

@article{PhysRevE.111.034120,
  title = {Nonequilibrium statistical mechanics revealed by Doob $h$ transform and variational autoregressive networks},
  author = {Zhao, Yixin and Tang, Ying and Zhang, Pan},
  journal = {Phys. Rev. E},
  volume = {111},
  issue = {3},
  pages = {034120},
  numpages = {15},
  year = {2025},
  month = {Mar},
  publisher = {American Physical Society},
  doi = {10.1103/PhysRevE.111.034120},
  url = {https://link.aps.org/doi/10.1103/PhysRevE.111.034120}
}

@article{tang2023neural,
  title={Neural-network solutions to stochastic reaction networks},
  author={Tang, Ying and Weng, Jiayu and Zhang, Pan},
  journal={Nat. Mach. Intell.},
  volume={5},
  number={4},
  pages={376--385},
  year={2023},
  publisher={Nature Publishing Group UK London},
  url = {https://www.nature.com/articles/s42256-023-00632-6}
}

@article{2lxs-wccj,
  title = {Space-Time Correlations in Monitored Kinetically Constrained Discrete-Time Quantum Dynamics},
  author = {Cech, Marcel and Cea, Mar\'{\i}a and Ba\~nuls, Mari Carmen and Lesanovsky, Igor and Carollo, Federico},
  journal = {Phys. Rev. Lett.},
  volume = {134},
  issue = {23},
  pages = {230403},
  numpages = {9},
  year = {2025},
  month = {Jun},
  publisher = {American Physical Society},
  doi = {10.1103/2lxs-wccj},
  url = {https://link.aps.org/doi/10.1103/2lxs-wccj}
}

@article{zhong2026scalable,
  title={Scalable Physics-Inspired Transformers for Spin Glasses},
  author={Zhong, Lu and Duan, Wenli and Liu, Jing and Zhang, Pan and Tang, Ying},
  journal={arXiv:2606.22984},
  year={2026},
  url={https://arxiv.org/abs/2606.22984}
}

@article{s6zj-vzdp,
  title = {Quantum Flow Matching},
  author = {Cui, Zidong and Zhang, Pan and Tang, Ying},
  journal = {PRX Intelligence},
  volume = {1},
  issue = {1},
  pages = {013009},
  numpages = {15},
  year = {2026},
  month = {Aug},
  publisher = {American Physical Society},
  doi = {10.1103/s6zj-vzdp},
  url = {https://link.aps.org/doi/10.1103/s6zj-vzdp}
}

@article{111_delRazo2026field,
  title = {Field theories and quantum methods for stochastic reaction-diffusion systems},
  author = {del Razo, Mauricio J. and Lamma, Tommaso and Merbis, Wout},
  journal = {Rev. Mod. Phys.},
  volume = {98},
  number = {1},
  pages = {015001},
  numpages = {44},
  year = {2026},
  month = {Jan},
  publisher = {American Physical Society},
  doi = {10.1103/9qlw-gyd7},
  url = {https://link.aps.org/doi/10.1103/9qlw-gyd7}
}

@article{RevModPhys.91.045002,
    title = {Machine learning and the physical sciences},
    author = {Carleo, Giuseppe and Cirac, Ignacio and Cranmer, Kyle and Daudet, Laurent and Schuld, Maria and Tishby, Naftali and Vogt-Maranto, Leslie and Zdeborov\'a, Lenka},
    journal = {Rev. Mod. Phys.},
    volume = {91},
    issue = {4},
    pages = {045002},
    numpages = {39},
    year = {2019},
    month = {Dec},
    publisher = {American Physical Society},
    doi = {10.1103/RevModPhys.91.045002},
    url = {https://link.aps.org/doi/10.1103/RevModPhys.91.045002}
}

@article{112_stinchcombe2012statistics,
  title = {Statistics of current-activity fluctuations in asymmetric flow with exclusion},
  author = {Stinchcombe, R. B. and de Queiroz, S. L. A.},
  journal = {Phys. Rev. E},
  volume = {85},
  number = {4},
  pages = {041111},
  year = {2012},
  month = {Apr},
  publisher = {American Physical Society},
  doi = {10.1103/PhysRevE.85.041111}
}

@article{113_dequeiroz2012current,
  title = {Current-activity versus local-current fluctuations in a driven flow with exclusion},
  author = {de Queiroz, S. L. A.},
  journal = {Phys. Rev. E},
  volume = {86},
  number = {4},
  pages = {041127},
  year = {2012},
  month = {Oct},
  publisher = {American Physical Society},
  doi = {10.1103/PhysRevE.86.041127}
}

@article{lin2026dynamical,
  title={Dynamical Partition Functions of Stochastic Dynamics via Variational Flows},
  author={Lin, Zequn and Tang, Ying},
  journal={arXiv:2606.10757},
  year={2026},
  url={https://arxiv.org/abs/2606.10757}
}

@article{gorissen2009density,
  title = {Density-matrix renormalization-group study of current and activity fluctuations near nonequilibrium phase transitions},
  author = {Gorissen, Mieke and Hooyberghs, Jef and Vanderzande, Carlo},
  journal = {Phys. Rev. E},
  volume = {79},
  number = {2},
  pages = {020101(R)},
  numpages = {4},
  year = {2009},
  month = {Feb},
  publisher = {American Physical Society},
  doi = {10.1103/PhysRevE.79.020101}
}

@article{lecomte2007continuous,
  title = {Thermodynamic formalism and large deviation functions in continuous time {Markov} dynamics},
  author = {Lecomte, Vivien and Appert-Rolland, C{\'e}cile and van Wijland, Fr{\'e}d{\'e}ric},
  journal = {C. R. Phys.},
  volume = {8},
  number = {5--6},
  pages = {609--619},
  year = {2007},
  publisher = {Elsevier},
  doi = {10.1016/j.crhy.2007.05.005}
}

@article{schutz1997exact,
  title = {Exact solution of the master equation for the asymmetric exclusion process},
  author = {Sch{\"u}tz, Gunter M.},
  journal = {J. Stat. Phys.},
  volume = {88},
  number = {1--2},
  pages = {427--445},
  year = {1997},
  month = {Jul},
  publisher = {Springer},
  doi = {10.1007/BF02508478}
}

@article{schutz1993phase,
  title = {Phase transitions in an exactly soluble one-dimensional exclusion process},
  author = {Sch{\"u}tz, Gunter and Domany, Eytan},
  journal = {J. Stat. Phys.},
  volume = {72},
  number = {1--2},
  pages = {277--296},
  year = {1993},
  month = {Jul},
  publisher = {Springer},
  doi = {10.1007/BF01048050}
}

@article{sinibaldi2026neural,
  title = {Time-Dependent Neural Galerkin Method for Quantum Dynamics},
  author = {Sinibaldi, Alessandro and Hendry, Douglas and Vicentini, Filippo and Carleo, Giuseppe},
  journal = {Phys. Rev. Lett.},
  volume = {136},
  pages = {120402},
  year = {2026},
  doi = {10.1103/kqvx-dl54}
}

@article{vandewalle2025manybody,
  title = {Many-body dynamics with explicitly time-dependent neural quantum states},
  author = {Van de Walle, Anka and Schmitt, Markus and Bohrdt, Annabelle},
  journal = {Mach. Learn.: Sci. Technol.},
  volume = {6},
  pages = {045011},
  year = {2025},
  doi = {10.1088/2632-2153/ae0f39}
}

\end{document}